\documentclass[aps,prd,twocolumn,english,nofootinbib]{revtex4-1}
\usepackage[T1]{fontenc}
\usepackage[latin9]{inputenc}
\usepackage{natbib}
\usepackage{color}
\usepackage{url}
\usepackage{babel}
\usepackage{array}
\usepackage{float}
\usepackage{amsmath}
\usepackage{ulem}
\usepackage{stmaryrd}
\usepackage{stackrel}
\usepackage{graphicx}
\usepackage{tikz}
\usetikzlibrary{shapes.geometric, arrows}

\tikzstyle{startstop} = [cylinder, rounded corners, minimum width=1.5cm, minimum height=0.5cm, text width=0.8cm, text centered, draw=black, fill=blue!30, shape border rotate=90]
\tikzstyle{io} = [trapezium, trapezium left angle=70, trapezium right angle=110, minimum width=4cm, minimum height=1cm, text width=2.5cm, text centered, draw=black, fill=blue!30]
\tikzstyle{process} = [rectangle, text width=3cm, minimum width=3cm, minimum height=1cm, text centered, draw=black, fill=yellow!30]
\tikzstyle{processl} = [rectangle, text width=3cm, minimum width=3cm, minimum height=1cm, text centered, draw=black, fill=green!30]
\tikzstyle{postprocess} = [rectangle, text width=3.8cm, minimum width=4cm, minimum height=1cm, text centered, draw=black, fill=yellow!30]
\tikzstyle{postprocessl} = [rectangle, text width=3.8cm, minimum width=4cm, minimum height=1cm, text centered, draw=black, fill=green!30]
\tikzstyle{exprocess} = [rectangle, text width=3cm, minimum width=3cm, minimum height=1cm, text centered, draw=black, fill=red!30]
\tikzstyle{exprocessl} = [rectangle, text width=3.8cm, minimum width=4cm, minimum height=1cm, text centered, draw=black, fill=red!30]
\tikzstyle{inprocess} = [rectangle, text width=2.5cm, minimum width=3cm, minimum height=1cm, text centered, draw=black, fill=green!30]
\tikzstyle{decision} = [diamond, minimum width=1.5cm, minimum height=0.7cm, text width=2.08cm, text centered, draw=black, fill=green!30]
\tikzstyle{arrow} = [thick, ->, >=stealth]
\tikzstyle{arrowboth} = [thick, <->, >=stealth]

\usepackage[unicode=true,pdfusetitle,
 bookmarks=true,bookmarksnumbered=false,bookmarksopen=false,
 breaklinks=false,pdfborder={0 0 1},backref=false,colorlinks=true]{hyperref}
\hypersetup{
 linkcolor=blue,urlcolor=blue,citecolor=blue}
\makeatletter
\providecommand{\tabularnewline}{\\}
\usepackage{blindtext}

\makeatother
\begin{document}
\title{Swarm-intelligent search for gravitational waves from eccentric binary
mergers}
\author{Souradeep Pal}
\email{sp19rs015@iiserkol.ac.in}

\affiliation{Indian Institute of Science Education and Research Kolkata, Mohanpur,
Nadia - 741246, West Bengal, India}
\author{K Rajesh Nayak}
\email{rajesh@iiserkol.ac.in}

\affiliation{Indian Institute of Science Education and Research Kolkata, Mohanpur,
Nadia - 741246, West Bengal, India}
\begin{abstract}
We implement an eccentric search for compact binary mergers based on particle swarm optimization. Orbital eccentricity is an invaluable input for understanding the formation scenarios of the binary mergers and can play a pivotal role in finding their electromagnetic counterparts. Current modelled searches rely on pre-computed template banks that are computationally expensive and resistant towards expanding the search parameter space dimensionality. On the other hand, particle swarm optimization offers a straightforward algorithm that dynamically selects template points while exploring an arbitrary dimensional parameter space. Through extensive evaluation using simulated signals from spin-aligned eccentric binary mergers, we discovered that the search exhibits a remarkable autonomy in capturing the effects of both eccentricity and spin. We describe our search pipeline and revisit some of the merger candidates from the gravitational wave transient catalogs.
\end{abstract}
\date{\today}

\maketitle

\section{Introduction}
Till the third observing run~(O3), the Advanced LIGOs and the Advanced
Virgo have collectively contributed to over 90 confident detections of transient gravitational wave (GW) events~\citep{abbott2019gwtc,abbott2021gwtc,abbott2021gwtc2_1,abbott2021gwtc3}. These include the first observations of binary black holes~(BBHs),
binary neutron stars~(BNSs), and the neutron star-black hole~(NSBH)
systems~\citep{abbott2016observation,abbott2017gw170817,abbott2021observation}.
In addition, several marginal candidate events have also been reported. Geographic separation of Virgo from the twin LIGO detectors has often aided the localization of the sources in the sky. The KAGRA and the GEO600 detectors also participated but their detection capabilities were marginal \citep{10.1093/ptep/ptac073}. With the addition of the LIGO-India observatory later in this decade, the network of second generation detectors should be complete~\citep{Saleem_2022}. 

Probing the orbital eccentricity of compact binary coalescences~(CBCs) is vital
for understanding their formation mechanism. Its accurate measurement on a
sizeable population of merger events can help to decide over the dominant
formation process, for example a dynamical capture or an isolated evolution.
The earlier understanding was that the eccentricity could radiate away quickly,
and the orbit becomes circularized before the signal enters the sensitive band
of the detectors. However, it may be possible that binary systems formed via a
dynamical interaction would not have enough time to circularize and can carry a
non-zero eccentricity. Although the LVK Collaboration have not reported eccentric
events in the past~\citep{abbott2019search}, there are several attempts
to find signatures of eccentricity in the LVK events~\citep{wu2020measuring,romero2019searching,gayathri2022eccentricity,Romero-Shaw_2021,Romero-Shaw_2022,Romero-Shaw_2020,10.1093/mnrasl/slaa084,10.1093/mnras/staa2120,PhysRevD.107.064024,gamba2023gw190521,iglesias2022reassessing}.
Currently, these efforts are not concerned with the detection stage but are usually
conducted as offline parameter estimation with schemes based on Bayesian techniques.
However, studies that infer binaries' formation channels based on the eccentric
parameter estimation of events flagged by non-eccentric searches could be biased.

The standard technique to search for CBC signals in the detector data involves cross-correlating the detector output with a bank of template waveforms that attempt to model the gravitational radiation. The current searches look for the spin effects in terms of the aligned or anti-aligned spins of the merger components, in addition to the component masses~\citep{messick2017analysis,adams2016low,usman2016pycbc,chu2022spiir}.
In principle, including extensive parameters in the templates describing the source
should benefit the search process. In practice, however, the benefit critically hinges on
the compuational cost and the feasibility of implementation. It is non-trivial to further
expand the parameter dimensionality of the current template-bank based searches.
However, if an inspiraling binary neutron star signal has a residual eccentricity
while entering the detector sensitive band, it would modulate the signal for
a large number of low frequency cycles before fading through
its late inspiral phase. Thus, a significant fraction of the signal-to-noise ratio
(SNR) can be accumulated from the low-frequency inspiral part, potentially improving
its significance and aiding the sky localization. However, the LVK collaboration
currently uses only the aligned-spin searches for generating candidates for their
follow-up. Also note that \texttt{BAYESTAR}, which computes the skymaps in low latencies,
is not optimized for localizing eccentric sources \citep{singer2016rapid}. 

An earlier search targeting non-spinning eccentric BNSs using a stochastic template placement algorithm could not lead to any interesting eccentric BNS candidate \citep{nitz2020search}. If eccentricity is associated with some of the detectable events, a generic eccentric search can enhance the significance of some marginal events flagged by the aligned-spin searches. After this paper describing spin-aligned eccentric searches was originally submitted, a similar search appeared for neutron star binaries but with a template bank that is about two orders of magnitude larger than a corresponding aligned-spin bank~\citep{dhurkunde2023search}. That search reports no new significant candidate from the third Observing Run of the Advanced LIGO and Advanced Virgo.

In this work, we demonstrate a spin-aligned eccentric search using the Particle
Swarm Optimization (PSO) algorithm. Previously, PSO based searches have been explored towards conducting aligned-spin and even fully precessing searches but never included eccentricity~\citep{PhysRevD.81.063002,srivastava2018toward,PhysRevD.101.082001}. Here, we include the orbital eccentricity ($\varepsilon$) as a search parameter in addition to the component masses and their aligned-spins. To generate template waveforms for BNS systems, we use the inspiral-only \texttt{TaylorF2Ecc} model. It is a fast, frequency-domain model supporting small eccentricities in the presence of aligned-spins \citep{moore2016gravitational}. For injecting signals from simulated sources, we primarily use the eccentric version of the time-domain \texttt{TEOBResumS} model (called \texttt{TEOBResumS-DALI}) that can incorporate high eccentricities in the presence of aligned-spins \citep{chiaramello2020faithful,nagar2021effective}. Here we also use it for generating templates for BBH systems. Currently, neither of these waveform models incorporate the mean anomaly parameter and thus, all our injections and the results are obtained at an arbitrary, fixed zero mean anomaly. The waveforms are computed at a reference frequency of 20 Hz. Our key observations and insights are summarized as follows.
\begin{enumerate}
\item  To explore a given search parameter space, the PSO algorithm iteratively computes the templates points on the fly while optimizing a given detection statistic. This approach is different from that of a template-bank which depends on a fixed, pre-determined set of template points. The PSO algorithm thus, avoids the \textit{minimal-match} criterion implemented in the template-bank searches but spends more time in exploring the region of the parameter space where the signal is more likely to be found. Some parts of the search parameter space generally require larger sampling than others. For example, the signal recovery in the low mass BNS region need more template points than that in the heavier BBHs. When data from more than one detector are available, the PSO algorithm can optimize a statistic computed over the detector network.
\item Matched-filtering of the data is preceded by template waveform generation in the PSO framework. Eccentric templates are in general costlier than the quasi-circular ones impacting the overall cost of the search. We find that the above strategy is effective in recovering the effects of eccentricity in the presence of the aligned-spins. However, we are currently limited by the parameter space coverage of the waveform models and the speed of template generation on-the-fly.
\item Eccentricity primarily modulates the inspiral component of the signals and the SNR gain is thus, mostly from the low frequency content of the signals. On the other end, truncating signals till $512$ Hz by downsampling reduces the cost of matched-filtering with little to no impact on recovered SNRs. If eccentricity is absent in the signals, the eccentric search approximately behaves like the usual aligned-spin search in terms of signal recovery.
\end{enumerate}
This article is organized as follows. Section \ref{sec:Search-Algorithm},
outlines the general search mechanism. First we describe the standard
PSO algorithm in locating the optimal template for a given segment
of data. We then implement a coincident search algorithm for finding
event candidates with multiple detetors. Finally we discuss how a
statistical significance is assigned to the coincident events. Section
\ref{sec:Eccentric-Injections} demonstrates its effectiveness on
simulated and real sources. We show the accuracy of recovering the
injected parameters and the SNRs using some agnostic populations of BNS
and BBH systems from coloured Gaussian noise. We discuss some challenges
in vetoing the noise artefacts from the real instruments. We also report the
re-analysis of a majority of gravitational wave transient catalog (GWTC) events with the proposed eccentric search. We finally include a parallelization scheme to reduce the runtime of the search in section \ref{subsec:Latency}.
%
%\begin{center}
\begin{figure*}
\begin{centering}
\includegraphics[scale=0.7]{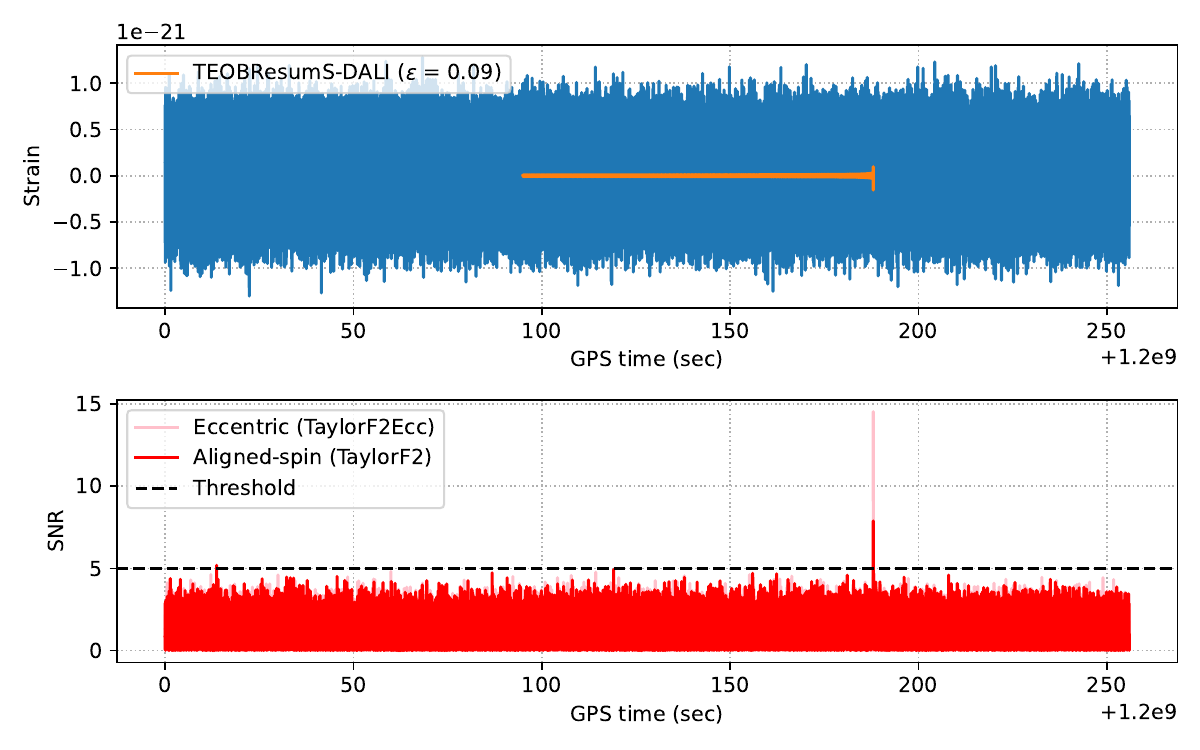}
\par\end{centering}
\caption{SNR timeseries for a typical eccentric injection recovered with the eccentric and the aligned-spin templates at the injected parameters; the injected signal is for a source with component masses $(1.4M_{\varodot},\,1.4M_{\varodot})$. The horizontal black dotted line marks the SNR threshold for generating triggers from individual detectors.~\label{fig:sig_snr}}
\end{figure*}
%\par\end{center}
\vspace{-0.2cm}
\section{Search Algorithm\label{sec:Search-Algorithm}}
\subsection{Detection principle\label{subsec:Detection-principle}}
In this section, we briefly review the overall method for coincident
detection of compact binary coalescences. The data recorded by a gravitational
wave detector is a timeseries of discretely sampled strain, $h(t)$ containing occasional gravitational signals in the detector's noise. The detection problem is to essentially identify instances of transient astrophysical signals in the timeseries data. This is achieved by cross-correlating the strain data with a modeled form of the astrophysical signal, called a \textit{template} waveform, $q(t,\, \boldsymbol{\theta})$. A template waveform depends on several parameters, $\boldsymbol{\theta}$ modelling the coalescing binary. The cross-correlation is performed in the frequency domain where the noise power spectral density (PSD), $S\textsubscript{n}(f)$ is used to down-weight the effect of noise, in a process called \textit{matched-filtering}. The output of matched-filtering is the signal-to-noise ratio, $\rho(t)$ defined as,
\begin{equation}
\rho^{2}(t;\boldsymbol{\theta})=4\Re{\int_{f\textsubscript{min}}^{f\textsubscript{max}}\frac{\tilde{h}(f)\tilde{q}^{*}(f;\boldsymbol{\theta})}{S\textsubscript{n}(f)}}e^{2\pi ift}df\label{eq:mfo}\,,
\end{equation}
where a tilde represents a function in the frequency domain. 
Whenever there is a close match between the true signal in the data
and a template waveform, the SNR timeseries attains a peak at the arrival time of the signal in the detector as shown in Fig.~\ref{fig:sig_snr}. However at times, detector noise can also mimic GW signals by causing sharp peaks in the SNR timeseries. These occur mainly due to the non-stationarity and non-Gaussian nature of the detector noise and are called \textit{glitches}. Glitches increase the chances of false alarms and affect the sensitivity to true astrophysical signals. The peaks in the SNR timeseries above a chosen threshold which arise due to the noise glitches and any potential astrophysical signals are considered as \textit{triggers}. The triggers undergo several consistency checks to veto out possible glitches. To get rid of some of the obvious loud noise transients, the signal-based power $\chi^{2}$ discriminator is used~\citep{allen2005chi}. Other vetoing techniques are also available for the noise triggers that dodge the standard consistency checks, especially from short duration templates \citep{nitz2018distinguishing}. The $\chi^{2}$ discriminators are designed to down-rank noise triggers in a quantity called the \textit{reweighted} SNR while preserving the original SNR values for the triggers with astrophysical origin~\citep{usman2016pycbc}. Remaining triggers are tested for time-coincidence with multiple detectors - the arrival times of astrophysical signals at any pair of detectors cannot differ by more than the time-of-flight for light between the detectors with some tolerance for the measurement errors. \textit{Events} that qualify these tests are assigned a statistical significance based on the rate of false alarms, which is described in section \ref{subsec:Significance-of-candidates}. The general detection problem is extensively discussed in the literature, for example~\citep{sathyaprakash2009physics,schutz1997introduction,PhysRevD.44.3819}.
\vspace{-0.5cm}
\subsection{Swarm-intelligence\label{subsec:Swarm-intelligence}}
%\begin{center}
\begin{figure*}
\begin{centering}
\includegraphics[scale=0.57]{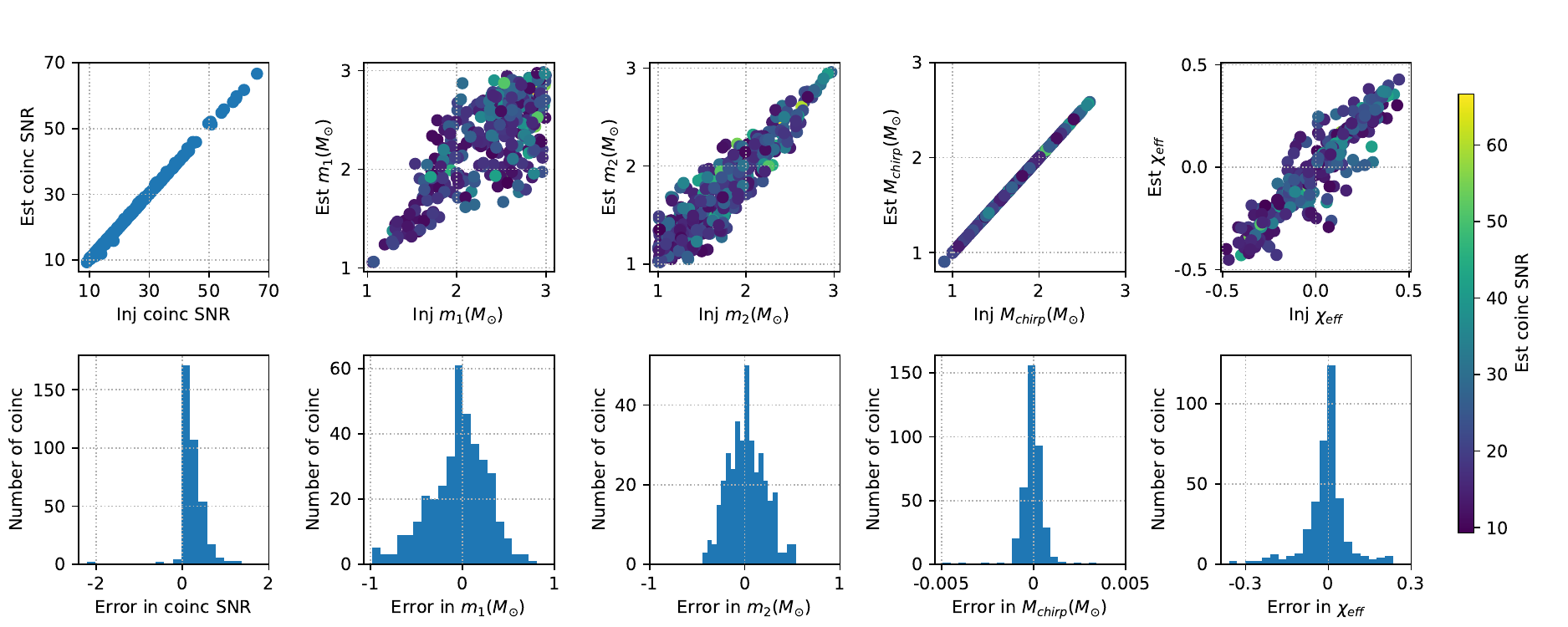}
\par\end{centering}
\caption{The accuracy in the recovery of the coincident SNR and the source parameters for the HL network with the proposed algorithm for non-eccentric BNS injections in Gaussian noise. The noise segments are coloured with the PSDs projected for the fourth Observing Run (O4) of the Advanced LIGO. Each point represents one injection and the colourmap indicates the recovered coincident SNR. The errors are obtained by subtracting the estimated values from the injected ones. Note that we have considered $m_{1} > m_{2}$ throughout this work. \label{fig:bns_acc}}
\end{figure*}
%\par\end{center}
Identifying a potential signal in a given segment of data hinges on finding a close match between the signal and a template waveform. As described in the previous section, one needs to optimize the function $\rho(t,\boldsymbol{\theta})$ as given in the Eq.~(\ref{eq:mfo}) over the desired range of parameters $\boldsymbol{\theta}$. In this section, we describe how the proposed algorithm locates the optimal template position which is represented by $\hat{\boldsymbol{\theta}}$. The algorithm begins with a fixed number of randomly distributed template points which we refer to as a \textit{swarm}. The individual members of the swarm, which are known as \textit{particles}, communicate and share the knowledge gained during the search process. Each particle with a position $x_{i}^{d}$ and a velocity $v_{i}^{d}$, where $i$ is the particle identifier and $d$ is the dimension index, constitutes a candidate for the optimal template $\hat{\boldsymbol{\theta}}$. The initial random distribution of the particles can be chosen to be uniform in each of the search dimensions. The role of the algorithm is to iteratively improve a given \textit{fitness} function by proposing a new set of positions to the particles and computing the fitness function at the newly proposed locations. The swarm eventually converges over few tens of iterations (typically less than 20) while the value of the fitness function saturates at an optimal value. Note that the fitness function must be carefully chosen since its repeated evaluation can be costly. For example, the simplest choice is the matched-filter SNR maximized over time for a single detector with Gaussian noise. However, a noise glitch can cause a very large SNR value and can deceive the algorithm. Thus we replace the SNR with the reweighted SNR for the network of detectors for maximization, as discussed in Section-\ref{subsec:Multiple-detectors}. 
%\begin{center}
%\par\end{center}
To take any subsequent step, a particle considers the following three factors-
\begin{enumerate}
\item[ (a)] the \textit{inertial} factor - the tendency to continue in the same direction as in the previous step,
\item[(b)] the \textit{cognitive} factor - the homesickness to go back to the best location achieved by itself, and
\item[(c)] the \textit{social} factor - the longing to join the best pool of peer achievers. 
\end{enumerate}
These three tendencies are reflected through the main three terms on the right-hand side of Eq.~\ref{eq:vel}. So after computing the fitnesses of the individual particles at any given step, the algorithm identifies two quantities: (a) the template point with the largest fitness function among all the particles, which is called \textit{global-best} or \textit{g-best}, denoted as $g^{d}$, and (b) the template point with the highest fitness achieved by the $i^{\rm th}$  particle till the current step, which is called \textit{personal-best} or \textit{p-best}, denoted as $p_{i}^{d}$. The overall velocity at $(t+1)^{\rm th}$ step can be expressed as: 
\begin{widetext}
\begin{equation}
v_{i}^{d}(t+1)=\alpha\;r_{\alpha}v_{i}^{d}(t)+\beta\;r_{\beta}[p_{i}^{d}(t)-x_{i}^{d}(t)]+\gamma\;r_{\gamma}[g^{d}(t)-x_{i}^{d}(t)]\label{eq:vel}\,,
\end{equation}
\end{widetext}
where $r_{\alpha},r_{\beta}$ and $r_{\gamma}$ are uniformly distributed
random numbers between $0$ and $1$. The constant weights $\alpha,\beta$
and $\gamma$ control the degree of convergence and exploration of the swarm. A PSO search is usually stable within a broad range of values for these \textit{hyperparameters}. A rough set of values is ($\alpha$, $\beta$, $\gamma$) = (0.7, 0.35, 0.4), but fine-tuning is possible depending upon the search. The positions of the particles are updated using the simple rule given below,
\begin{equation}
x_{i}^{d}(t+1)=x_{i}^{d}(t)+v_{i}^{d}(t+1)\label{eq:pos}\,.
\end{equation}
\begin{figure}[t]
\begin{centering}
\includegraphics[scale=0.6]{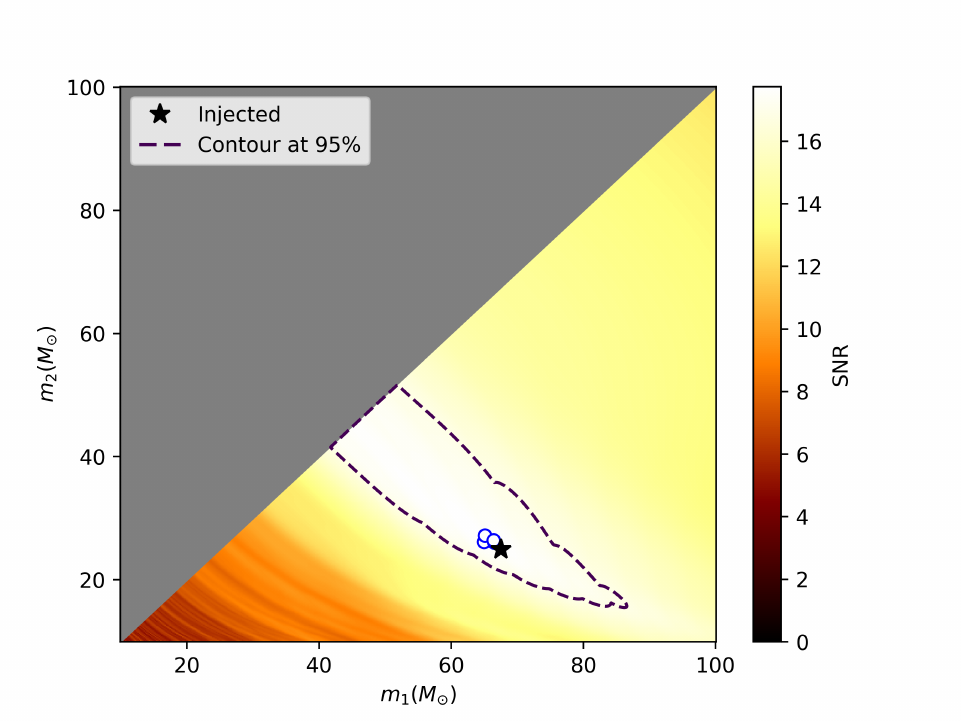}
\par\end{centering}
\caption{The fitness function has a broad set of values with several nearby peaks which can confuse the convergence to the true parameter values. The blue rings show the g-best locations finally explored by a few independent swarms- all lie on the \textit{ridge} while attaining very close fitness function values. The dashed contour contains the nearly \textit{degenerate} region in the template space where fitness function is greater than 0.95 times that of the injected value.
\label{fig:ridge}}
\end{figure}

As an initial test of robustness of the PSO algorithm, we demonstrate the recovery of simulated BNS sources from a small population of about 500 injections in Gaussian noise. These are non-eccentric BNS injections with moderate aligned-spin components; the component masses are randomly drawn from $[1M_{\varodot},3M_{\varodot}]$. To locate the optimal template parameters for the HL detector network, we used 5k particles and allowed up to 15 steps, aggregated to about 75k template evaluations and that many matched-filter computations per detector per injection. The initial distribution of the particles are uniform for the search dimensions given by
\begin{equation}
\boldsymbol{\theta}=\{m\textsubscript{1}, m\textsubscript{2}, \chi\textsubscript{1z}, \chi\textsubscript{2z}\}\label{eq:aspin}\,.
\end{equation}
This corresponds to an aligned-spin search using PSO. A summary of the simulation results are shown in Fig.~\ref{fig:bns_acc}. The injected coincident SNR is the coincident SNR obtained from the template with the injected parameters. Here an average loss of about 0.4\% is observed in the recovered coincident SNR from that of the injected values. We also observed that to tackle less dense regions of the search space, like that of the NSBH and BBHs, requires even lesser sampling of the parameter space in general. The final g-best location in the parameter space represents an optimal solution with the obtained template parameters $\hat{\boldsymbol{\theta}}$. Here the point estimates of the parameters describing the simulated sources are reasonably unbiased, notably the chirp mass, $M\textsubscript{chirp}$ and the effective spin parameter, $\chi\textsubscript{eff}$ as shown in Fig.~\ref{fig:bns_acc}. The modifications to this algorithm to adapt to the real instrumental noise are described in the next subsection.
\vspace{-0.2cm}
\subsection{Multiple detector optimization\label{subsec:Multiple-detectors}}
%\begin{center}
\begin{figure*}[t]
\begin{centering}
\begin{tikzpicture}[node distance=1.8cm]

\footnotesize{}

\node (h1data) [startstop] {H1 strain};
\node (l1data) [startstop, right of=h1data, xshift=2.65cm] {L1 strain};
\node (v1data) [startstop, right of=l1data, xshift=2.65cm] {V1 strain};

\node (process1) [process, below of=h1data, yshift=-0.0cm] {Downsample to 1024 Hz, high pass filter from 15 Hz and local PSD generation};
\node (process2) [process, below of=l1data, yshift=-0.0cm] {Downsample to 1024 Hz, high pass filter from 15 Hz and local PSD generation};
\node (process3) [process, below of=v1data, yshift=-0.0cm] {Downsample to 1024 Hz, high pass filter from 15 Hz and local PSD generation};

\node (process4) [exprocess, below of=process1, yshift=-0.0cm] {Matched-filtering with the common set of templates};
\node (process5) [exprocess, below of=process2, yshift=-0.0cm] {Matched-filtering with the common set of templates};
\node (process6) [exprocess, below of=process3, yshift=-0.0cm] {Matched-filtering with the common set of templates};

\node (process7) [process, below of=process4, yshift=-0.3cm] {Compute triggers with SNR $>$ 4, rank them using the $\chi^{2}$-test, discard triggers with newSNR/SNR $ < 0.75$};
\node (process8) [process, below of=process5, yshift=-0.3cm] {Compute triggers with SNR $>$ 4, rank them using the $\chi^{2}$-test, discard triggers with newSNR/SNR $ < 0.75$};
\node (process9) [process, below of=process6, yshift=-0.3cm] {Compute triggers with SNR $>$ 4, rank them using the $\chi^{2}$-test, discard triggers with newSNR/SNR $ < 0.75$};

\node (process10) [postprocessl, below of=process8, yshift=-0.8cm] {Pick the trigger with highest newSNR/SNR per detector and combine to get the joint fitness function for each particle};

\node (processpso) [processl, below of=process10, yshift=-0.1cm] {Compute pbest, gbest and mean swarm radius};

\node (decision1) [decision, below of=processpso, yshift=-1cm] {Meets termination condition?};

\node (postprocess) [postprocessl, below of=decision1, yshift=-1.4cm] {Generate foreground coincidences, update trigger buffers and compute background events};

\node (process11) [postprocessl, left of=decision1, xshift=-5.5cm] {Update the position and the velocity of the particles and apply boundary conditions};

\node (process12) [exprocessl, left of=process10, xshift=-5.5cm] {Compute templates at the position of the particles (assume a random distribution for iteration-1)};
%%%%%%%%%%%%%%%%%%%%%%%%%%%%%%%%%%%%%%%%%%%%
%%%%%%%%%%%%%%%%%%%%%%%%%%%%%%%%%%%%%%%%%%%%
\draw [arrow] (h1data) -- (process1);
\draw [arrow] (l1data) -- (process2);
\draw [arrow] (v1data) -- (process3);

\draw [arrow] (process1) -- (process4);
\draw [arrow] (process2) -- (process5);
\draw [arrow] (process3) -- (process6);

\draw [arrow] (process4) -- (process7);
\draw [arrow] (process5) -- (process8);
\draw [arrow] (process6) -- (process9);

\draw [arrow] (process7) -- (process10);
\draw [arrow] (process8) -- (process10);
\draw [arrow] (process9) -- (process10);

\draw [arrow] (process10) -- (processpso);

\draw [arrow] (processpso) -- (decision1);

\draw [arrow] (decision1) -- node[anchor=south] {No} (process11);

\draw [arrow] (decision1) -- node[anchor=east] {Yes} (postprocess);

\draw [arrow] (process11) -- (process12);

\draw [arrow] (process12) |- (process4);

\draw [arrow] (process12) |- (process4);

\draw [arrowboth] (process4) -- (process5);

\draw [arrowboth] (process5) -- (process6);

\end{tikzpicture}

\thispagestyle{empty}

%\end{document}
\par\end{centering}
\caption{Flowchart showing steps involved in the PSO search with network optimization. The boxes in light red are computationally the most expensive steps, followed by the yellow
and the green ones. \label{fig:flow_chart}}
\end{figure*}
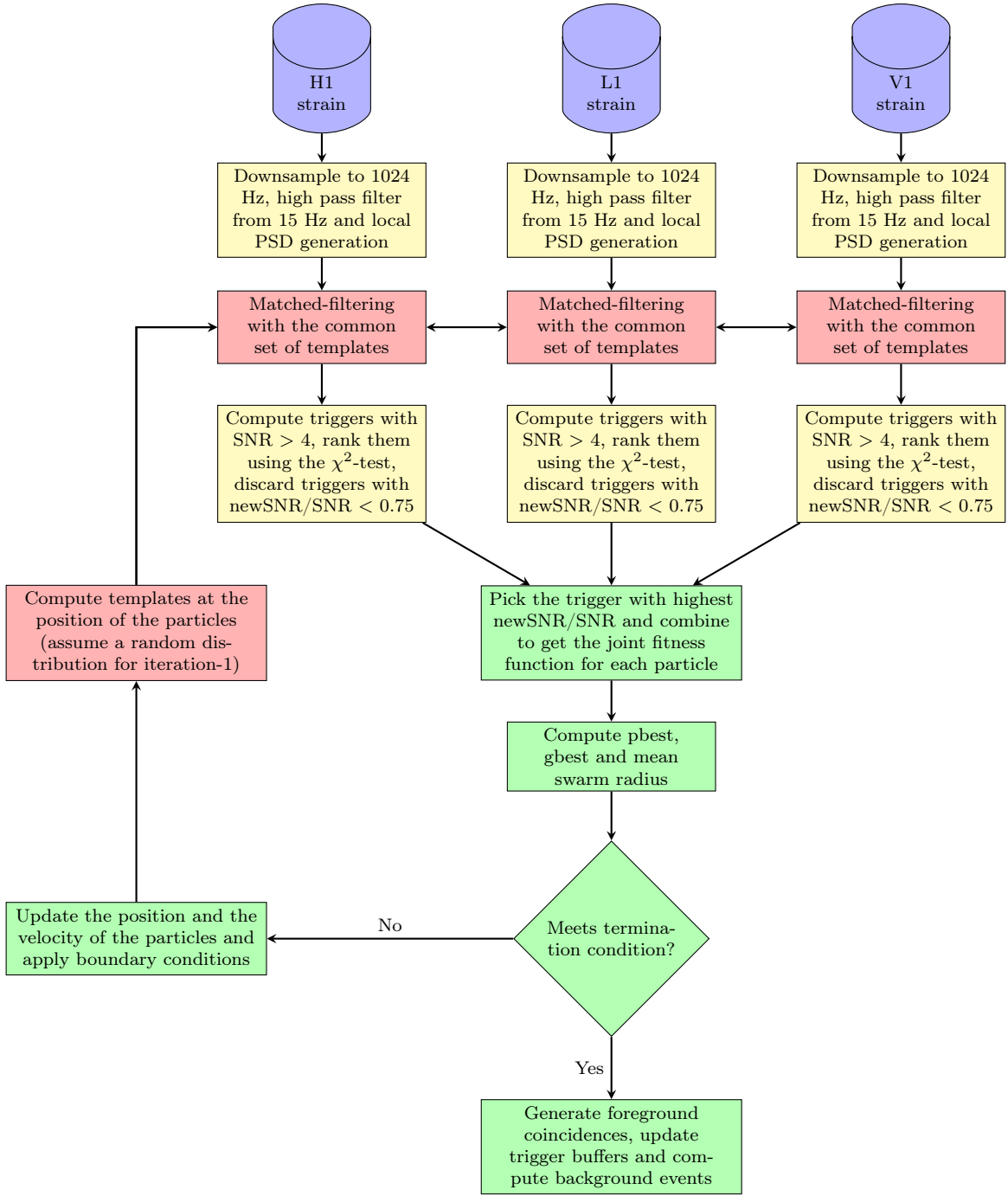
%\par\end{center}
When searching in the \textit{coincident} mode, the processes described earlier in section \ref{subsec:Swarm-intelligence} can, in principle, proceed independently and in parallel for all the available detectors. However, in such a case, the templates are also computed independently for each detector and the point of convergence can differ due to several reasons. The noise contributions from the individual detectors can slightly shift the convergence point. Otherwise, several local peaks form an elongated band of almost \textit{degenerate} templates with similar SNRs as shown in Fig.~\ref{fig:ridge}, which can confuse the convergence. But a true gravitational wave signal should have the times of arrival and the source parameters consistent across all the available detectors. To ensure the second criterion, the same set of template parameters should optimize a fitness function computed for the detector network. Besides, for a given template, a \textit{stretch} of data can contain several noise triggers, however, such triggers are expected to be uncorrelated in time and appear at just one detector. A data stretch here represents the data recorded by the available detectors within a fixed GPS start and end times. Thus, to minimize spurious optimization of glitch triggers, a natural choice is to jointly optimize the reweighted SNRs across the network of detectors. The \textit{joint} fitness function for a network of $N$ detectors is given by,
\begin{equation}
\tilde{\rho}=\sqrt{\stackrel[j=0]{N}{\sum}\tilde{\rho}_{j}^{2}}\label{eq:net_snr}\ ,
\end{equation}
where $\tilde{\rho}_{j}$ is the reweighted SNR of the trigger most likely arising from an astrophysical signal at the $j$th detector. Template points that trigger to a loud glitch in one of the detectors are downranked by the veto processes and the swarm further inhibits convergence to such template points over the detector network. An additional advantage of the above procedure is that a single set of computation of the template waveforms serves the entire network of detectors. It is especially useful when generating the template waveforms is expensive as is generally the case in an eccentric search. A similar strategy is also employed to optimize the network SNR of the candidates found in the \texttt{PyCBC Live} search \citep{Canton_2021}.

The steps involved in the optimization process are described as follows. These are also summarized in Fig.~\ref{fig:flow_chart}. To demonstrate the process, we assume that the coincident strain data from different detectors exist in the computer storage. The data blocks for each detector are divided into segments overlapping with their immediate neighbours in time to account for the corrupted ends of the SNR timeseries due to filtering artifacts. The duration of the data stretch is chosen in a way that an in-band inspiral signal lies completely within a data stretch. The PSO algorithm explores each data stretch one after another. Data segments are downsampled till 1024 Hz as discussed later in Section \ref{sec:Eccentric-Injections}. PSDs used in the matched-filters are computed from the segments using Welch's method~\citep{1161901}.

To evaluate the fitness function for a given data stretch at any template point, the template waveform at $x_{i}$ is constructed and cross-correlated with the stretch of data. This results into an SNR timeseries per detector whose ends are discarded from the either side. Triggers from the remaining portion of the SNR timeseries above a chosen threshold, typically $> 4$, are examined. We compute the reweighted SNR values at the trigger points using the power $\chi^{2}$-test. Additionally, we compute the ratio of the reweighted SNR and the SNR for the triggers and use it to rank them for their astrophysical origin. Triggers with a ratio larger than $0.75$ (referred to as \textit{primary} triggers) are retained. The highest ranked trigger, i.e. the one with the highest ratio, from each detector is used to update the fitness function associated with a given particle as given in (Eq. \ref{eq:net_snr}). Note that we do not consider triggers with a ratio less than $0.75$. If no such trigger exists for a given template point for the data stretch, the corresponding particle proceeds for optimization with its previous value of the fitness function.

Based on the updated values of the fitness function of all the particles, the p-best and the g-best locations and fitnesses are computed as described in subsection ~\ref{subsec:Swarm-intelligence}. Following this, Eq. \ref{eq:vel}-\ref{eq:pos} are used to assign a new set of positions to the particles for the next iteration. The algorithm also checks whether any particle has drifted beyond the desired search boundaries- a \textit{reflecting} boundary puts back an outlier inside in a symmetric position while flipping its velocity~\footnote{While this is suitable in most cases, we find that an \textit{adsorbing}, $\varepsilon=0$ boundary is more effective for the eccentric dimension towards the recovery of non-eccentric sources, where instead of putting the outlier back to a symmetric location, we put it on the boundary.}. The algorithm additionally checks for a terminal condition of a convergence below a small mean radius of the swarm or a preset timeout, whichever is earlier. The boundary conditions can improve the termination and also ensures no hang-up occurs due to non-availability of template parameters outside the region allowed by a waveform model. Note that these steps, namely evolving the position-velocity, applying the boundary conditions and the termination checks are computationally trivial. At this point, the current iteration is complete. 

The algorithm repeats this procedure for few tens of iterations and returns the final g-best template point for the data stretch. We further compute the triggers from the optimized g-best template (or the \textit{detection template}) in a similar fashion as described earlier and keep all the triggers that survive the ratio cut. These triggers from the detection template (referred to as \textit{secondary} triggers) are used to form coincidences in a post-processing step as described in the next subsection.
\vspace{-0.2cm}
%%%%%
\subsection{Significance of candidates\label{subsec:Significance-of-candidates}}
%%%%
\begin{figure}[b]
\begin{centering}
\includegraphics[scale=0.7]{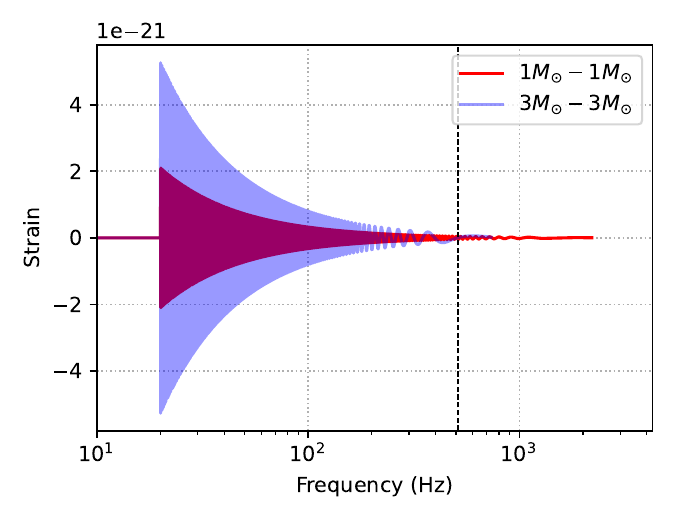} 
\par\end{centering}
\caption{Frequency-domain waveforms generated using the \texttt{TaylorF2Ecc}
model for $\varepsilon=0.1$. The black dashed line represents the high
frequency cutoff at 512 Hz. This cutoff leads to only negligible loss in the SNRs of low mass binaries near $(1M_{\varodot},1M_{\varodot})$ as described in the text but roughly reduces the cost of the matched-filtering to one half of that using a standard 1024 Hz cutoff. \label{fig:downsample}}
\end{figure}
\begin{figure*}[t]
\begin{centering}
\includegraphics[scale=0.55]{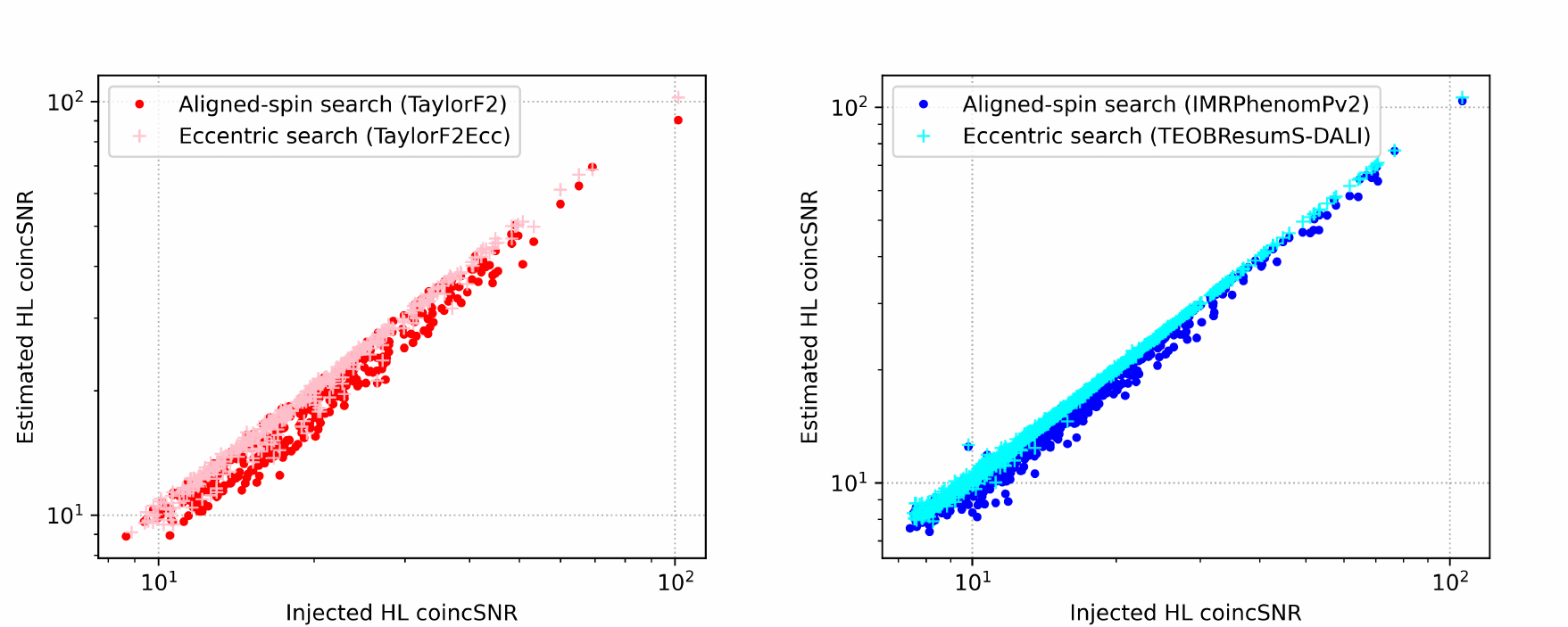}
\par\end{centering}
\caption{Recovering the HL coincident SNRs for spin-aligned eccentric systems: BNS injections (left) and BBH injections (right) in O4 projected Gaussian noise. Both the eccentric (plus) and the aligned-spin (dot) searches are performed on identically prepared datasets.\label{fig:SNR_Accuracy}}
\end{figure*}
When the iterations finish, the algorithm returns the detection template for a given stretch of data with the associated triggers from the individual detectors. The detection template point is already consistent in the source parameters across all the detectors. Now the secondary triggers are tested for any zero-lag time coincidence to create \textit{foreground} events. A time coincidence test requires triggers from different detectors to occur within a time window defined by the light's travel time between the detectors plus 5 ms accounting for any possible timing measurement errors. Several such foreground candidates, which pass both the parameter and the time coincidence tests, may still not be of astrophysical origin. To assess the possibility of a noise origin, one assigns a statistical significance to a foreground candidate and/or a probability of being of astrophysical origin, $p\textsubscript{astro}$ \citep{farr2015counting,capano2017systematic}. Here we assign a \textit{false alarm rate} (FAR) to the foreground events using the method of time-shifting of triggers obtained from processing the data immediately available near (or before) the candidate~\citep{wkas2009background}. The algorithm maintains a sufficient buffer of the cumulative secondary triggers resulting from the detection templates of the neighbouring (or past) data stretches. Note that the different stretches can have their optimized g-best template points located throughout the parameter space. Thus, though the secondary triggers for a given data stretch have consistency over the source parameters, the collection of such triggers over multiple stretches do not enforce the source parameter consistency~\footnote{An appropriate strategy to utilize even the primary triggers for the background estimates in the PSO search can be explored in a future study. However, the cumulative triggers from the entire space are generally more prone to picking up abrupt noise and can even degrade the background statistic.}. A test of accuracy of this FAR estimation strategy with real data is described in subsection~\ref{subsec:Instrument-noise}.

If the triggers obtained from a pair of detectors are artificially shifted with respect to each other by a time more than travel time of light between the two detectors, then the resulting coincidences cannot be of astrophysical origin. Such artificial \textit{background} events are also ranked by their artificial network statistic value and a FAR estimate is assigned to a foreground candidate as
\begin{equation}
FAR\sim\frac{N_{\geq}}{T_{bg}}.\label{eq:far_eq}
\end{equation}
Here $N_{\geq}$ is the number of background events with a network statistic equal to or greater than the foreground event. $T_{bg}$ is computed as the artificial time over which the background coincident events are manufactured. In the simple case of a two detector network, it can be approximated by $T\textsubscript{live}^2 / T\textsubscript{slide}$, where $T\textsubscript{live}$ is the coincident live time of the two detectors and $T\textsubscript{slide}$ is a non-astrophysical time-slide duration. To compute a multi-detector background, the triggers from the most sensitive detector may be slided with remaining detectors fixed~\citep{davies2020extending}. Even considering triggers from Virgo, a time-slide duration $T\textsubscript{slide}$ of 50 ms can be safely chosen and as many artificial events are created as possible. When additional detectors are ready with similar sensitivities, a general approach with each detector sliding may be appropriate. So, to attempt a significant confidence of FAR less than 1 per 100 years to a given foreground event, a buffer of a sufficient duration of recent triggers must be kept which turns out to be a minimum of 3.5 hrs in our case. Normally, a large window for buffering triggers is more exposed to less recent and abrupt changes in the detector noise. Online LVK searches, however, require a higher FAR threshold for considering candidates for follow-up processes. In this work, we have mostly used only 4096 seconds to reach FARs upto 1 per 10 years.
\begin{figure}[b]
\begin{centering}
\includegraphics[scale=0.7]{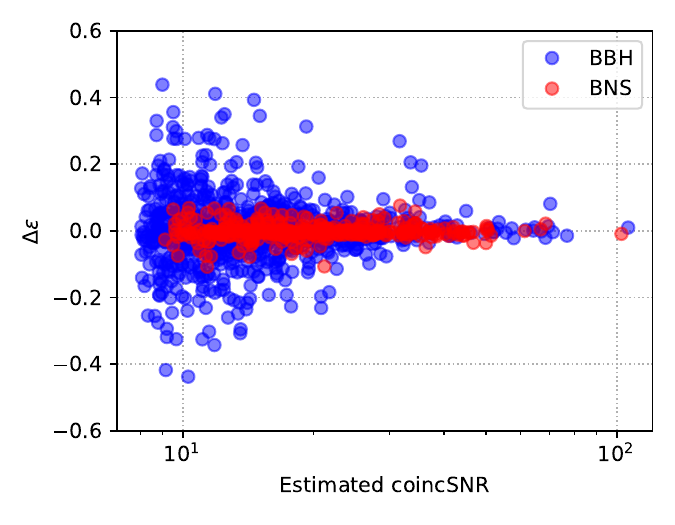} 
\par\end{centering}
\caption{A plot of the error in the eccentricity- the difference in the value estimated by the g-best template from that of the injected value plotted against the coincident
SNRs. \label{fig:ecc_err}}
\end{figure}
\vspace{-4.4pt}
%%%%%
\section{Eccentric Search\label{sec:Eccentric-Injections}}
%%%%%
\vspace{0.4cm}
\begin{figure*}[htbp]
\centering
\includegraphics[scale=0.7]{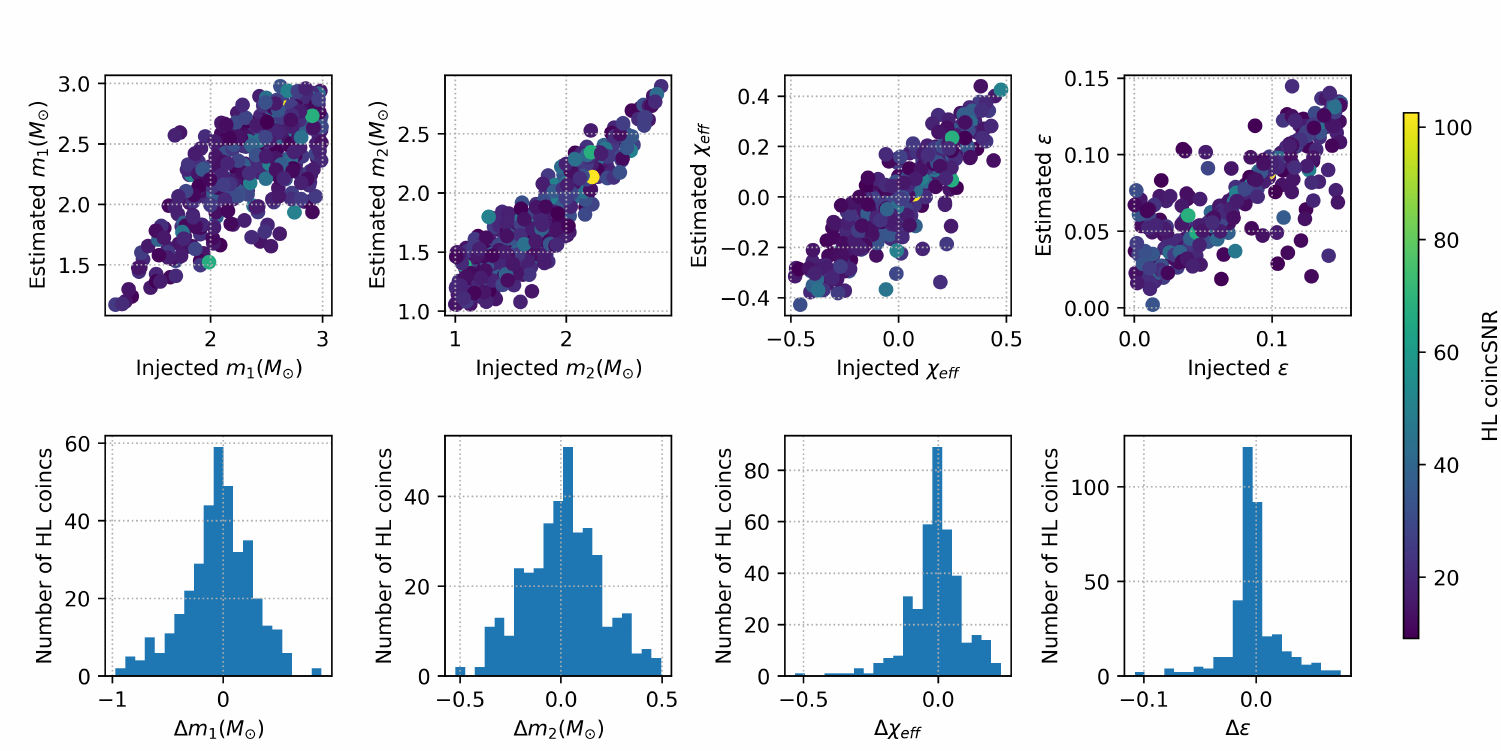} 
\caption{The PSO algorithm is sensitive to small eccentricities in the presence of aligned spins for BNS systems at around 75k matched-filtering operations per detector per injection. The \texttt{TaylorF2Ecc} waveform model supports only small eccentricities for comparable mass binaries. \label{fig:bns_chirp}}
\smallskip
\includegraphics[scale=0.7]{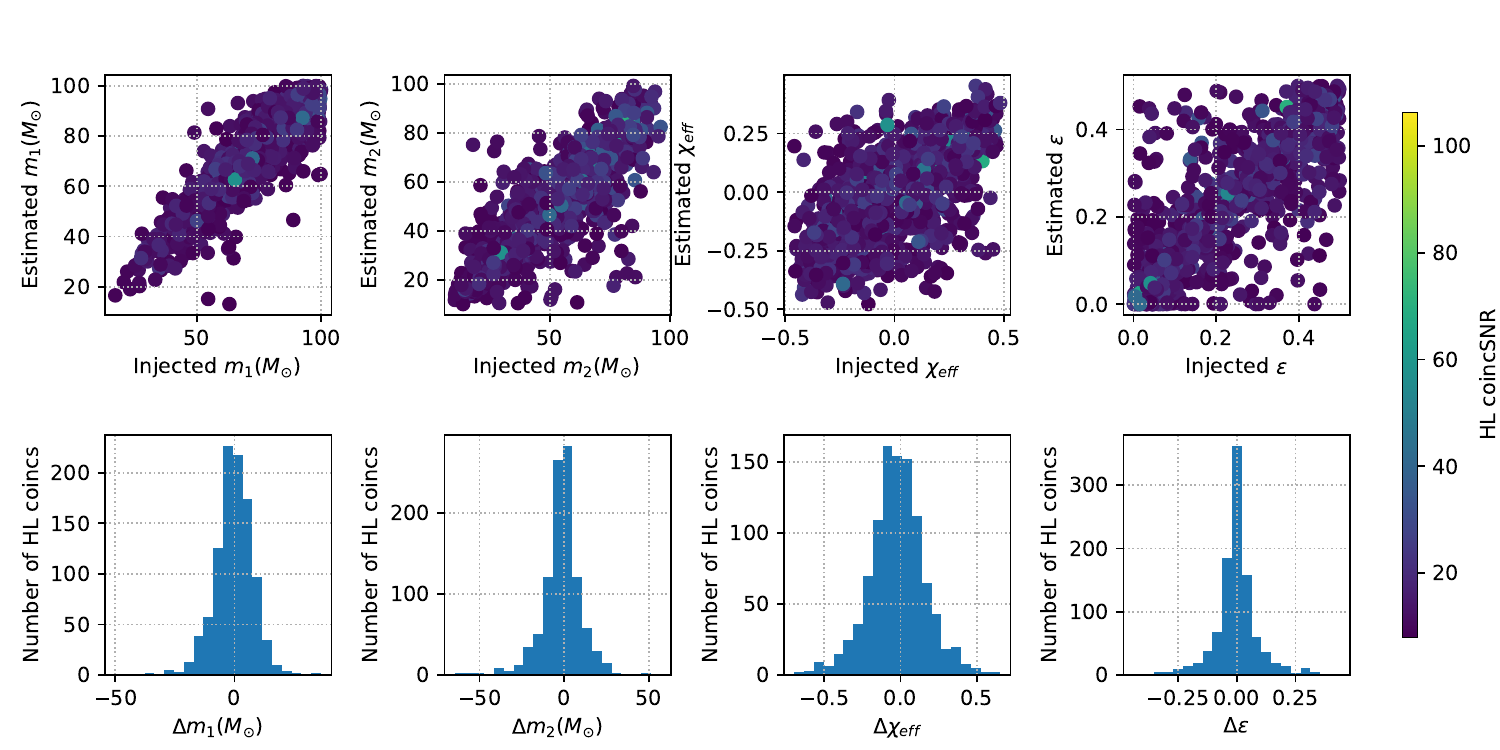} 
\caption{Approximately bias free point estimates are available
for spin-aligned eccentric BBH systems using the time-domain \texttt{TEOBResumS-DALI} model. Note that the points lying far off the diagonal usually have low SNR values as indicated in Fig.~\ref{fig:ecc_err}. \label{fig:BBH_para}}
\end{figure*}
\vspace{-0.4cm}
\begin{figure*}[t]
\begin{centering}
\includegraphics[scale=0.85]{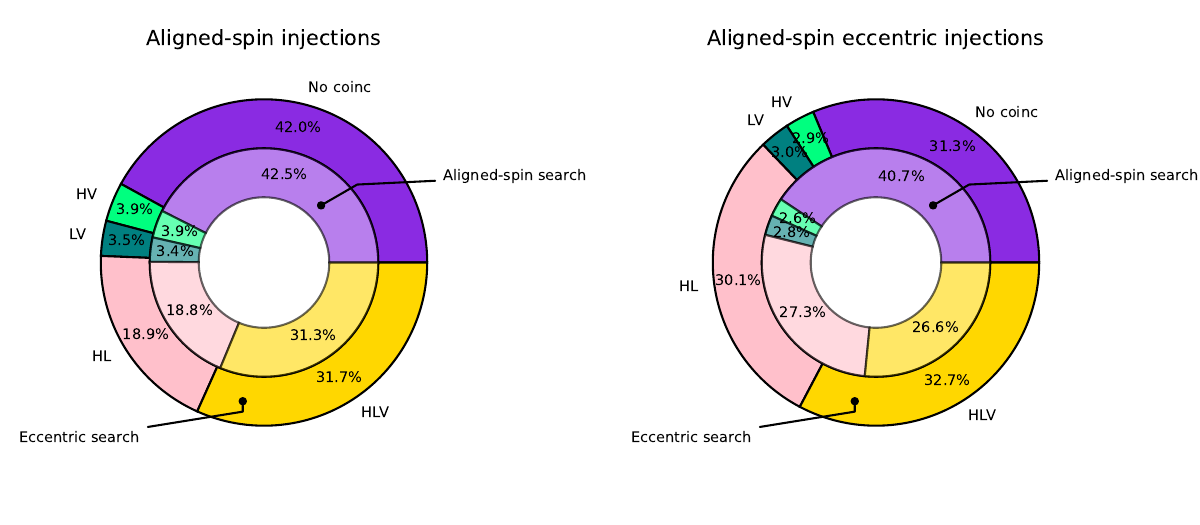} 
\par\end{centering}
\vspace{-12pt}
\caption{Summary of the simulated searches on spin-aligned non-eccentric BBH injections using \texttt{IMRPhenomPv2} (left) and on spin-aligned eccentric injections using \texttt{TEOBResumS-DALI} (right). Note that the recovery with the eccentric search (outer rings) is at par with that of the aligned-spin search (inner rings) for aligned-spin non-eccentric injections, whereas it recovers significantly more coincidences for aligned-spin eccentric injections (right). Fraction of triple coincidences also increase primarily because some of the subthreshold Virgo signals now clear the trigger threshold due to improved SNR with the eccentric search.\label{fig:pie_chart-1}}
\end{figure*}
In the previous section, we described a general framework to search for CBC signals using the PSO algorithm. Now, we demonstrate the algorithm specifically designed for spin-aligned eccentric sources. Here the search dimensions are given by 
\begin{equation}
\boldsymbol{\theta}=\{m\textsubscript{1}, m\textsubscript{2}, \chi\textsubscript{1z}, \chi\textsubscript{2z}, \varepsilon\}\label{eq:espin}\,.
\end{equation}
We use a uniform initial distribution of particles in the added eccentric dimension, $\varepsilon$. In the absence of the occasional and loud disturbances, or glitches, the real instrument data can be considered as nearly Gaussian. Thus in this section, we first demonstrate the recovery of simulated signals with our algorithm in coloured Gaussian noise. The signals are projected from random sky locations onto a network of simulated detectors and then injected into the noise segments. Here we at most use the 3-detector combination, from H1,\,L1, and V1, and use the publicly available O4 representative PSDs to colour the noise segments \citep{pubpsd}. We then discuss the response of the search to the real data from the instruments when no likely signals are present. For each set of results obtained with the aligned-spin eccentric search using PSO, we also show results from a corresponding aligned-spin non-eccentric search with PSO.

To reduce the cost of matched-filtering, data are downsampled to 1024 Hz while remaining sensitive to almost the full inspiral part of the signal as shown in Fig.~\ref{fig:downsample}. Note that the \texttt{TaylorF2Ecc} is an inspiral-only model and does not anyway contain much power in the high frequencies. The loss in the recovered SNR from such a high frequency cutoff would be just around 1\% in the case of a perfectly matching template for a $(1M_{\varodot},1M_{\varodot})$ binary and even smaller for heavier binaries. We currently set the low frequency cutoff at 20 Hz. Since, the effects of eccentricity are more in the lower frequencies, it is reasonable to expect improvements with a lower low frequency cutoff.
%%%%%%%%%%%%%%
\subsection{Eccentric BNSs\label{subsec:Eccentric-BNSs}}
%%%%%%%%%%%%%%%%%%
In this section, we generate mock signals from a simulated population of about 500 eccentric BNS sources using the \texttt{TEOBResumS-DALI} model and inject them into the Gaussian noise simulated for the two LIGO detectors. Agnostic about the true astrophysical distribution, the merger component masses are uniformly drawn from $[1M_{\varodot},3M_{\varodot}]$, with moderately large values of the aligned-spin components in $[-0.5,0.5]$ and small eccentricities from $[0.0,0.15]$. We use \texttt{TaylorF2Ecc} to recover the signals using the PSO algorithm. The \texttt{TaylorF2Ecc} waveform model basically draws the parameter space coverage from the widely used \texttt{TaylorF2} model to search for BNS systems. For long duration signals, we expect to see an insignificant SNR loss due to the missing merger-ringdown parts in the template waveforms as shown in~(Fig. \ref{fig:SNR_Accuracy}). We test the accuracy of the
PSO algorithm to mainly recover the coincident SNR and the injected
binary parameters. Note that our primary goal here is to demonstrate
the effectiveness of the PSO algorithm, but it also tests the agreement
of the waveform parameters between the two models. The choice of
the injection distribution may not be astrophysically well informed, however, the luminosity distances to the simulated sources are chosen from $[20,150]$ Mpc
distributed uniformly in volume, giving a realistic range of network
SNRs for O4 and beyond. In Fig.~\ref{fig:bns_chirp}, the accuracy in recovering the injected parameters is plotted for the component masses, the effective spin parameters and the eccentricity. We find that the PSO algorithm is readily sensitive to these small eccentrities with the total 75k matched-filtering operations per detector per injection. Note that, even a small eccentricity can modulate the signal over a large number of cycles, attributing to
the improvement in the recovered {\it coincident} SNR as shown in the 
Fig.~\ref{fig:SNR_Accuracy}. It also may be concluded that the waveform parameter values are largely precise for the two models from different formalisms. Note that whatever deviation observed here also includes the fact that the templates do not span the merger-ringdown phases of the signals.
%%%%%%%%%%%%%%%
\subsection{Eccentric BBHs\label{subsec:Eccentric-BBHs}}
%%%%%%%%%%%%%%%%%
\vspace{-0.5cm}
\begin{center}
\begin{figure*}[t]
\begin{centering}
\includegraphics[scale=0.55]{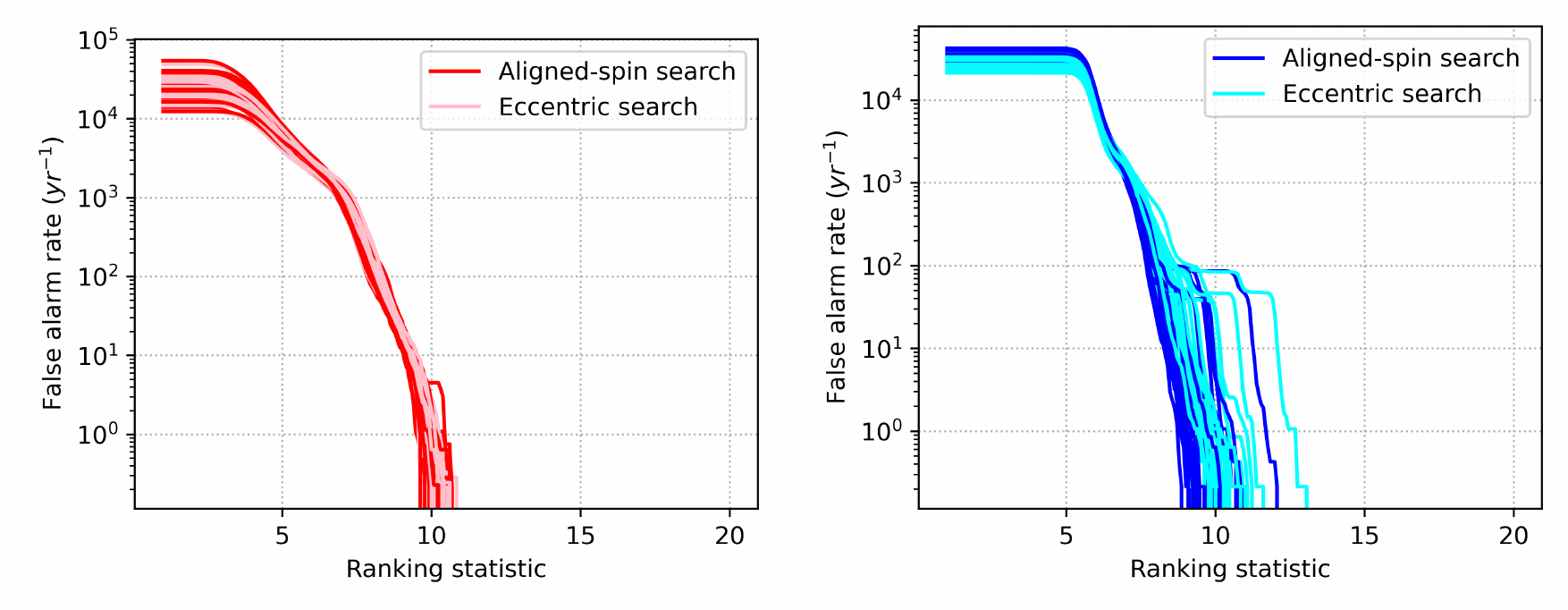}
\par\end{centering}
\centering{}\caption{Rate of coincident background events plotted against their ranking statistic for the aligned-spin vs the eccentric search for some 25 random blocks (4096 seconds) of strain data from O3 (left BNS and right BBH). Qualitatively, we observe no significant difference in the background event rates with the additional eccentric dimension in the search. We note that slight variations are possible given the stochastic nature of the search algorithm. Also, no significant foreground candidate is obtained from these datasets. However, fewer step-like curves, common to both the BBH searches, apparently indicate the presence of few louder triggers than expected whose origin we do not explore here. \label{fig:dim4/5}}
\end{figure*}
\par\end{center}
Here we use the \texttt{TEOBResumS-DALI} waveform model for both the signal injection and the template generation. This waveform model contains the inspiral-merger-ringdown phases required for the SNR build up, especially for the high mass binaries. As before, Fig.~\ref{fig:BBH_para} summarizes the recovered parameters for the HL coincidences out of 1000 injections in the mass range $[10M_{\varodot},100M_{\varodot}]$, aligned-spins within a range of $[-0.5,0.5]$ and eccentricities up to $0.5$. The luminosity distances of the sources are chosen uniformly in volume between 500-3000 Mpc. The search was performed using about 30k matched-filtering operations per detector per injection. 

To further demonstrate capabilities of the eccentric search, we perform a simulation with a large population of about 10k BBH injections. We prepare two injection sets with identical source parameters except eccentricity as follows- (1) non-eccentric aligned-spin BBH injections using \texttt{IMRPhenomPv2}, and (2) eccentric aligned-spin BBHs using \texttt{TEOBResumS-DALI}. In this investigation, we use the HLV detector network with the projected O4 model sensitivities for the Advanced LIGO and the Advanced Virgo.

We simultaneously conduct the following two searches on both of the injection sets- both having identical configurations but (1) with eccentric spin-aligned templates using the \texttt{TEOBResumS-DALI} model, and (2) with non-eccentric spin-aligned templates using the \texttt{IMRPhenomPv2} model. Since, the data segments are synthetically gaussian and hence do not contain noise glitches, we consider injections as {\it found}, once the triggers cross an SNR threshold of 5 in coincidence with at least two detectors. The search results are summarized in Fig.~\ref{fig:pie_chart-1}. We quickly note the following about the eccentric search. 
\begin{itemize}
\item 
The recovery rate is notably greater- an overall of $\sim9.1\%$ more coincident events are recovered than that of the aligned-spin search. 
\item 
The overall \textit{quality} of recovery is better~(e.g. the network SNRs) in comparison with the aligned-spin recovery for the commonly found injections (however, not explicit from the figure). 
\item 
A fraction of extra events feature in the triple (HLV) coincidences with the eccentric search, indicating potential improvements their significance and the follow-ups. 
\item 
Both the searches recover similar number and fractions of the total coincidences in the case of spin-aligned non-eccentric injections. 
\end{itemize}
Note that these statistics can vary depending on the choice of the injection population, detector PSDs and the instrument noise with glitches playing a key role in the false alarm rate associated with an event. We discuss some of these challenges in the next section.
%%%%%%%%%
\subsection{Background event rates\label{subsec:Instrument-noise}}
%%%%%%%%%%
\vspace{-1.5cm}
\begin{center}
\begin{figure}[b]
\includegraphics[scale=0.7]{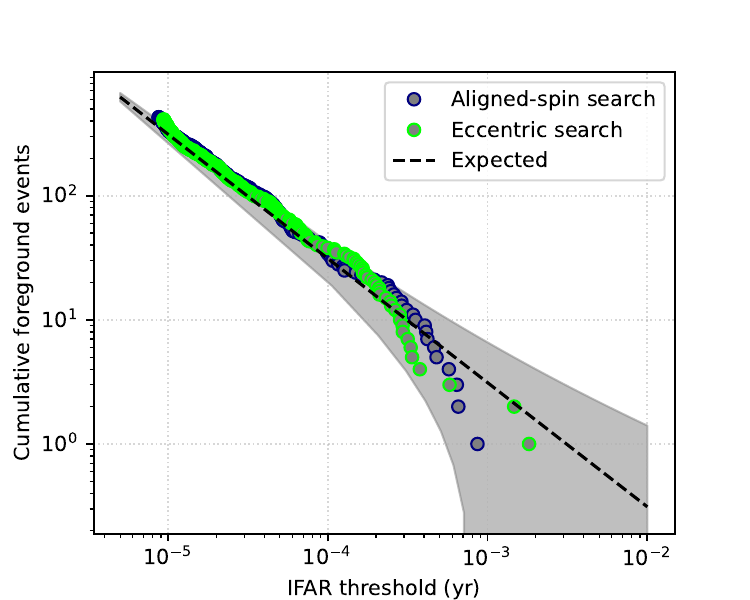}
\caption{The cumulative number of foreground events obtained from the real datasets plotted against the inverse FAR threshold. The dashed line represents the expected cumulative number of such events above a given IFAR value. The grey-shaded region indicates the 95\% Poisson uncertainty region.\label{fig:IFARcombine}}
\end{figure}
\par\end{center}
\begin{center}
\small
\begin{table*}
\scriptsize
\setlength\extrarowheight{1.5pt}
\begin{centering}
\begin{tabular}{>{\centering}p{2.3cm}>{\centering}p{1.2cm}>{\centering}p{2cm}>{\centering}p{1.2cm}>{\centering}p{1.2cm}>{\centering}p{1cm}>{\centering}p{1cm}>{\centering}p{1cm}>{\centering}p{1cm}>{\centering}p{1cm}>{\centering}p{1cm}>{\centering}p{1cm}>{\centering}p{1cm}}
\hline 
{ Event} & { Triggered} & { GPS time} & { $m_{1}$} & { $m_{2}$} & { $\chi\textsubscript{eff}$} & { $\varepsilon$} & { FAR} & { $\tilde{\rho}$} & \multicolumn{4}{c}{{ $\rho$}}\tabularnewline
\cline{10-13} \cline{11-13} \cline{12-13} \cline{13-13} 
 &  & { (seconds)} & { $(M_{\varodot})$} & { $(M_{\varodot})$} &  &  & { (year$^{-1}$)} &  & { \texttt{PSO}} & { \texttt{PyCBC}} & { \texttt{MBTA}} & { \texttt{GstLAL}}\tabularnewline
\hline 
\hline 
{ GW150914} & { HL} & { 1126259462.4} & { 46.1} & { 26.3} & { 0.002} & { 0.073} & { 0.38} & { 23.7} & { 23.7} & { 23.6} & { -} & { 24.4}\tabularnewline
{ GW151012} & { HL} & { 1128678900.4} & { 26.6} & { 17.6} & { 0.099} & { 0.283} & { 0.38} & { 9.7} & { 9.8} & { 9.5} & { -} & { 10}\tabularnewline
{ GW170104} & { HL} & { 1167559936.5} & { 40.4} & { 20.7} & { -0.034} & { 0.035} & { 0.38} & { 13.1} & { 13.2} & { 13} & { -} & { 13}\tabularnewline
{ GW170729} & { HL} & { 1185389807.3} & { 70.1} & { 29.1} & { 0.051} & { 0.325} & { 2.26} & { 9.5} & { 9.9} & { 9.8} &  & { 10.8}\tabularnewline
{ GW170809} & { HL} & { 1186302519.7} & { 34.7} & { 27.2} & { -0.186} & { 0.277} & { 0.38} & { 10.2} & { 10.8} & { 12.2} & { -} & { 12.4}\tabularnewline
{ GW170814} & { HL} & { 1186741861.5} & { 37.5} & { 26.9} & { 0.179} & { 0.2} & { 0.38} & { 12.7} & { 15.0} & { 16.3} & { -} & { 15.9}\tabularnewline
{ GW170817{$^@$}} & { HL} & { 1187008882.4} & { 1.66} & { 1.15} & { -0.001} & { 0.01} & { 0.38} & { 32.4} & { 32.5} & { 30.9} & { -} & { 33.0}\tabularnewline
{ GW170818} & { HL} & { 1187058327.1} & { 44.8} & { 22.1} & { -0.185} & { 0.218} & { 0.38} & { 9.9} & { 10.0} & { -} & { -} & { 11.3}\tabularnewline
{ GW170823} & { HL} & { 1187529256.5} & { 51.2} & { 43.8} & { 0.264} & { 0.056} & { 0.38} & { 11.3} & { 11.6} & { 11.1} & { -} & { 11.5}\tabularnewline
{ GW190403\_051519} & { -} & { -} & { -} & { -} & { -} & { -} & { -} & { -} & { -} & { -} & { -} & { -}\tabularnewline
{ GW190408\_181802} & { HL} & { 1238782700.3} & { 41.9} & { 17.2} & { 0.002} & { 0.104} & { 0.38} & { 14.1} & { 14.1} & { 13.1} & { 14.4} & { 14.7}\tabularnewline
{ GW190413\_052954} & { HL} & { 1239168612.5} & { 54.9} & { 43.6} & { 0.126} & { 0.071} & { 0.38} & { 8.7} & { 8.7} & { 8.5} & { -} & { -}\tabularnewline
{ GW190413\_134308} & { HL} & { 1239198206.7} & { 82.6} & { 37.6} & { -0.297} & { 0} & { 0.38} & { 9.2} & { 9.5} & { 9.3} & { 10.3} & { 10.1}\tabularnewline
{ GW190421\_213856} & { HL} & { 1239917954.2} & { 63.5} & { 40.8} & { -0.148} & { 0.108} & { 0.38} & { 10.2} & { 10.2} & { 10.1} & { 9.7} & { 10.5}\tabularnewline
{ GW190503\_185404} & { HL} & { 1240944862.3} & { 49.4} & { 34.6} & { -0.097} & { 0.204} & { 0.38} & { 12.2} & { 12.2} & { 12.2} & { 12.8} & { 12}\tabularnewline
{ GW190512\_180714} & { HL} & { 1241719652.4} & { 27.5} & { 15.9} & { -0.068} & { 0.12} & { 0.38} & { 11.4} & { 11.6} & { 12.4} & { 11.7} & { 12.2}\tabularnewline
{ GW190513\_205428} & { -} & { -} & { -} & { -} & { -} & { -} & { -} & { -} & { -} & { 11.6} & { 13} & { 12.3}\tabularnewline
{ GW190514\_065416} & { -} & { -} & { -} & { -} & { -} & { -} & { -} & { -} & { -} & { -} & { -} & { 8.4}\tabularnewline
{ GW190517\_055101} & { HL} & { 1242107479.8} & { 54.4} & { 24.9} & { 0.365} & { 0.306} & { 0.38} & { 9.5} & { 9.5} & { 10.4} & { 10.8} & { 11.3}\tabularnewline
{ GW190519\_153544} & { HL} & { 1242315362.4} & { 92.6} & { 85.3} & { -0.499} & { 0.191} & { 2.63} & { 11.3} & { 11.3} & { 13.2} & { 13.7} & { 12.4}\tabularnewline
{ GW190521\_030229{$^*$}} & { HL} & { 1242442967.4} & { 143.5} & { 133.8} & { 0.223} & { 0.246} & { 0.009} & { 13.5} & { 13.7} & { 13.7} & { 13} & { 13.3}\tabularnewline
{ GW190521\_074359} & { HL} & { 1242459857.4} & { 51.8} & { 43.2} & { 0.162} & { 0.005} & { 0.38} & { 24.2} & { 24.4} & { 24} & { 22.2} & { 24.4}\tabularnewline
{ GW190527\_092055} & { HL} & { 1242984073.8} & { 60.7} & { 25.1} & { 0.088} & { 0.322} & { 0.38} & { 9.0} & { 9.0} & { -} & { -} & { 8.7}\tabularnewline
{ GW190602\_175927} & { HL} & { 1243533585.1} & { 93.1} & { 72.0} & { 0.036} & { 0.195} & { 0.38} & { 12.2} & { 12.2} & { 11.9} & { 12.6} & { 12.3}\tabularnewline
{ GW190620\_030421} & { -} & { -} & { -} & { -} & { -} & { -} & { -} & { -} & { -} & { -} & { -} & { 10.9}\tabularnewline
{ GW190630\_185205} & { -} & { -} & { -} & { -.} & { -} & { -} & { -} & { -} & { -} & { -} & { 15.2} & { -}\tabularnewline
{ GW190701\_203306} & { HLV} & { 1246048404.6} & { 69.9} & { 56.3} & { -0.235} & { 0.436} & { 2.63} & { 12.3} & { 12.3} & { 11.9} & { 11.3} & { 11.7}\tabularnewline
{ GW190706\_222641} & { HL} & { 1246487219.3} & { 90.3} & { 48.7} & { 0.148} & { 0.497} & { 0.38} & { 13.0} & { 13.0} & { 11.7} & { 11.9} & { 12.5}\tabularnewline
{ GW190708\_232457} & { -} & { -} & { -} & { -} & { -} & { -} & { -} & { -} & { -} & { -} & { -} & { 13.1}\tabularnewline
{ GW190719\_215514} & { HL} & { 1247608532.9} & { 53.9} & { 30.6} & { 0.184} & { 0.164} & { 0.38} & { 8.5} & { 8.5} & { -} & { -} & { -}\tabularnewline
{ GW190727\_060333} & { HL} & { 1248242631.9} & { 53.3} & { 48.5} & { 0.151} & { 0.007} & { 0.38} & { 11.7} & { 11.7} & { 11.4} & { 12} & { 12.1}\tabularnewline
{ GW190731\_140936} & { HL} & { 1248617394.6} & { 66.1} & { 16.7} & { -0.356} & { 0.09} & { 0.38} & { 8.6} & { 8.6} & { -} & { 9.1} & { 8.5}\tabularnewline
{ GW190803\_022701} & { HL} & { 1248834439.8} & { 58.6} & { 36.7} & { -0.119} & { 0.292} & { 5.64} & { 8.2} & { 8.5} & { 8.7} & { 9} & { 9.1}\tabularnewline
{ GW190805\_211137} & { HL} & { 1249074715.3} & { 88.5} & { 35.2} & { 0.326} & { 0.335} & { 0.38} & { 7.9} & { 7.9} & { -} & { -} & { -}\tabularnewline
{ GW190828\_063405} & { HL} & { 1251009263.7} & { 39.6} & { 36.1} & { 0.169} & { 0.164} & { 0.38} & { 15.7} & { 15.7} & { 13.9} & { 15.2} & { 16.3}\tabularnewline
{ GW190828\_065509} & { HL} & { 1251010527.8} & { 29.7} & { 13.9} & { 0.075} & { 0.158} & { 0.75} & { 11.0} & { 11.0} & { 10.5} & { 11.1} & { 10.8}\tabularnewline
{ GW190910\_112807} & { -} & { -} & { -} & { -} & { -} & { -} & { -} & { -} & { -} & { -} & { -} & { 13.4}\tabularnewline
{ GW190915\_235702} & { HL} & { 1252627040.6} & { 42.2} & { 28.5} & { -0.116} & { 0.175} & { 0.38} & { 12.8} & { 12.8} & { 13} & { 12.7} & { 13}\tabularnewline
{ GW190916\_200658} & { HL} & { 1252699636.8} & { 48.9} & { 35.9} & { 0.05} & { 0.259} & { 381.12} & { 6.7} & { 6.8} & { -} & { 8.2} & { 8.2}\tabularnewline
{ GW190925\_232845} & { HV} & { 1253489343.1} & { 28.2} & { 14.8} & { 0} & { 0.127} & { 0.38} & { 9.9} & { 9.9} & { 9} & { 9.4} & { -}\tabularnewline
{ GW190926\_050336} & { HL} & { 1253509434.0} & { 87.5} & { 16.2} & { -0.208} & { 0.338} & { 1.13} & { 8.1} & { 8.1} & { -} & { -} & { 9}\tabularnewline
{ GW190929\_012149} & { HL} & { 1253755327.4} & { 88.3} & { 46.9} & { -0.062} & { 0.14} & { 0.75} & { 9.6} & { 9.6} & { 9.4} & { 10.3} & { 10.1}\tabularnewline
{ GW191109\_010717} & { HL} & { 1257296855.2} & { 80.4} & { 59.6} & { 0.176} & { 0.333} & { 0.38} & { 16.0} & { 16.0} & { 13.2} & { 15.2} & { 15.8}\tabularnewline
{ GW191127\_050227} & { HL} & { 1258866165.5} & { 46.7} & { 26.5} & { -0.214} & { 0} & { 1.50} & { 8.7} & { 8.7} & { 9.5} & { 9.8} & { 10.3}\tabularnewline
{ GW191204\_110529} & { HL} & { 1259492747.5} & { 35.5} & { 27.6} & { 0.131} & { 0.023} & { 0.38} & { 9.6} & { 9.6} & { 8.9} & { 8.1} & { 9}\tabularnewline
{ GW191215\_223052} & { HL} & { 1260484270.3} & { 39.5} & { 21.1} & { 0.048} & { 0.085} & { 0.38} & { 10.6} & { 10.6} & { 10.3} & { 10.8} & { 10.9}\tabularnewline
{ GW191222\_033537} & { HL} & { 1261020955.1} & { 68.5} & { 51.6} & { 0.018} & { 0.092} & { 0.38} & { 11.4} & { 11.8} & { 11.5} & { 10.8} & { 12}\tabularnewline
{ GW191230\_180458} & { HL} & { 1261764316.4} & { 90.4} & { 37.9} & { 0.075} & { 0.109} & { 0.38} & { 9.8} & { 9.8} & { 9.6} & { 9.8} & { 10.3}\tabularnewline
{ GW200112\_155838} & { -} & { -} & { -} & { -} & { -} & { -} & { -} & { -} & { -} & { -} & { -} & { 17.6}\tabularnewline
{ GW200128\_022011} & { HL} & { 1264213229.9} & { 70.6} & { 52.2} & { 0.25} & { 0.078} & { 0.38} & { 10.1} & { 10.1} & { 9.8} & { 9.4} & { 10.1}\tabularnewline
{ GW200129\_065458} & { HLV} & { 1264316116.4} & { 40.4} & { 31.4} & { 0.156} & { 0.201} & { 0.38} & { 20.8} & { 25.5} & { 16.3} & { -} & { 26.5}\tabularnewline
{ GW200208\_130117} & { HLV} & { 1265202095.9} & { 51.7} & { 33.5} & { -0.13} & { 0.036} & { 0.38} & { 10.8} & { 10.8} & { 9.6} & { 10.4} & { 10.7}\tabularnewline
{ GW200208\_222617} & { -} & { -} & { -} & { -} & { -} & { -} & { -} & { -} & { -} & { -} & { 8.9} & { 8.2}\tabularnewline
{ GW200209\_085452} & { HL} & { 1265273710.1} & { 67.8} & { 37.1} & { -0.137} & { 0.149} & { 0.38} & { 9.0} & { 9.0} & { 9.2} & { 9.7} & { 10}\tabularnewline
{ GW200216\_220804} & { HL} & { 1265926102.8} & { 79.5} & { 32.7} & { 0.107} & { 0.12} & { 0.38} & { 8.8} & { 8.8} & { 9} & { 8.8} & { 9.4}\tabularnewline
{ GW200219\_094415} & { HL} & { 1266140673.1} & { 63.7} & { 39.0} & { -0.042} & { 0} & { 1.13} & { 10.0} & { 10.0} & { 9.9} & { 10.6} & { 10.7}\tabularnewline
{ GW200220\_061928} & { -} & { -} & { -} & { -} & { -} & { -} & { -} & { -} & { -} & { -} & { -} & { -}\tabularnewline
{ GW200220\_124850} & { -} & { -} & { -} & { -} & { -} & { -} & { -} & { -} & { -} & { -} & { 8.2} & { 8.2}\tabularnewline
{ GW200224\_222234} & { HLV} & { 1266618172.4} & { 49.6} & { 36.8} & { 0.105} & { 0.169} & { 0.38} & { 14.8} & { 18.1} & { 19.2} & { 19} & { 18.9}\tabularnewline
{ GW200225\_060421} & { HL} & { 1266645879.3} & { 23.0} & { 18.7} & { -0.037} & { 0.088} & { 0.38} & { 11.1} & { 12.3} & { 12.3} & { 12.5} & { 12.9}\tabularnewline
{ GW200302\_015811} & { -} & { -} & { -} & { -} & { -} & { -} & { -} & { -} & { -} & { -} & { -} & { 10.6}\tabularnewline
{ GW200306\_093714} & { -} & { -} & { -} & { -} & { -} & { -} & { -} & { -} & { -} & { 7.8} & { 8.5} & { -}\tabularnewline
{ GW200308\_173609} & { -} & { -} & { -} & { -} & { -} & { -} & { -} & { -} & { -} & { 7.9} & { 8.3} & { 8.1}\tabularnewline
{ GW200311\_115853} & { HLV} & { 1267963151.3} & { 40.7} & { 32.2} & { -0.019} & { 0.046} & { 0.38} & { 15.8} & { 17.1} & { 17} & { 16.5} & { 17.7}\tabularnewline
{ GW200322\_091133} & { -} & { -} & { -} & { -} & { -} & { -} & { -} & { -} & { -} & { 8} & { 9} & { -}\tabularnewline
\hline 
\end{tabular}
\par\end{centering}
\normalsize
\caption{Re-analysis of the BBH events in the range $[10M_{\varodot},100M_{\varodot}]$ from the GWTCs \citep{abbott2019gwtc,abbott2021gwtc,abbott2021gwtc2_1,abbott2021gwtc3} from the perspective of the eccentric BBH search. $^@$But we have also included \texttt{GW170817}, which is the only BNS event observed with at least two detectors. For comparison, the SNRs recovered by the LVK analyses are also given as reported in \citep{gwoscgwtc}. \texttt{PyCBC} results shown here are for the \texttt{PyCBC-broad} search. The event with ({$^*$}) has been re-analyzed as described in the text. \label{tab:gwtc_table}}
\end{table*}
\par\end{center}
\begin{figure*}[t]
\includegraphics[scale=0.55]{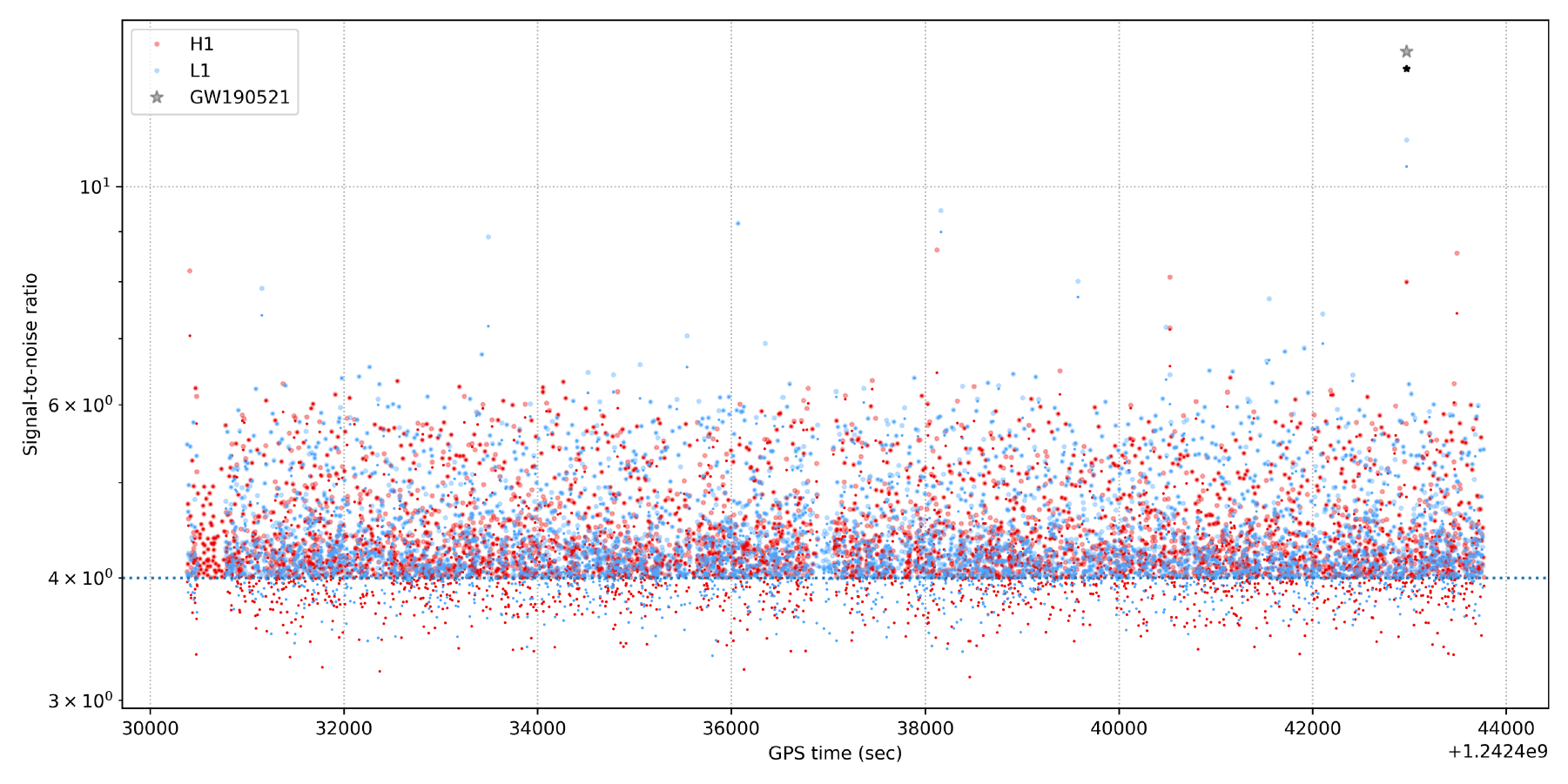}
\caption{Eccentric re-searching of \texttt{GW190521}. The data points shown here represent the secondary triggers resulting from various detection templates in their respective data stretches around the time of \texttt{GW190521}. The lighter (larger) blobs represent individual detector trigger SNRs and the darker (smaller) dots represent their reweighted SNR values, collected over a buffer time of about 3.5 hours before the event time. Background events generated from the triggers during this period in the HL network result into a false alarm rate of less than 1 per 100 years. Note that a coincident SNR value of the event is estimated to be 13.7 (shown in grey star) which is larger than any other background event in this case. \label{fig:GW190521}}
\end{figure*}
In the previous sections, we have shown that the eccentric searches
can outperform the aligned-spin searches in the presence of Gaussian noise. However, the data from a real detector can contain nearby noise glitches which can overwhelm the searches. To investigate this, we start with an hypothesis that there is no \textit{strong} astrophysical signal and data segments contain only detector noise and glitches. We perform the searches independently using the same configuration as described in section \ref{sec:Eccentric-Injections} on several data blocks from the third observing run (O3a) of the Advanced LIGOs not containing any obvious CBC candidates \citep{abbott2023open,RICHABBOTT2021100658}. For simplicity we use data from only two detectors H1 and L1. Each 4096 seconds data block is divided into segments of 256 seconds with a 50\% overlap, equally shared with its neighbouring segments and the analysis procedure described in Sec. \ref{subsec:Multiple-detectors} is followed. Thus each PSO instance effectively searches 128 seconds of data from the available detectors. Note that the triggers crossing an SNR threshold undergo the same signal consistency tests and the network reweighted SNR is used as the ranking statistic. For comparison, we also conduct the aligned-spin non-eccentric search using PSO on the same data segments with identical set-up. The results are summarized in Fig.~\ref{fig:dim4/5}.

While searching for the BNS systems, we find that both the searches
generate a similar rate of background events. The power $\chi^{2}$-test
is effective in this low mass region with long inspirals and vetoes
out most of the glitches. On the other hand, the eccentric BBH search uses a time domain model for the templates, where the filter artefacts can lead to higher rate of background events if not conditioned properly. We constrain the waveform with the tapering scheme for time-domain templates illustrated in \citep{mckechan2010tapering}. In addition, it is known that some of the glitches, such as the \textit{blips}, can successfully dodge the power $\chi^{2}$-test for the higher-mass templates. In the BBH searches, the component masses can go beyond 100 $M_{\varodot}$ and eccentricity can further shorten the template durations. Sine-gaussians can approximately model some of these glitches and separate them from GW signals by taking into account of the power at frequencies higher than those in the templates \citep{nitz2018distinguishing}. To benefit and extend the eccentric BBH search for high mass systems, this can be incorporated in the future when we expand the search space coverage.

However, to assess the accuracy of the estimation of the false alarm rates described earlier in subsection~\ref{subsec:Significance-of-candidates}, we use the triggers obtained from analyzing the above datasets. The cumulative number of all the foreground candidates are plotted against a given inverse false alarm rate (IFAR) threshold in Fig.~\ref{fig:IFARcombine}. Here the FARs are assigned using the combined triggers from both BNS and BBH searches, inclusive of those forming the foreground candidates. The expected number of noise events for a given FAR threshold over a coincident observation time is calculated using Eq.~\ref{eq:far_eq}. The figure demonstrates a general self consistency of the FARs assigned to any of the foreground candidates and indicates that the candidates are consistent with noise origin in this case.
%%%%%%%%%
\subsection{Eccentric re-searching of GWTC events\label{subsec:Eccentric-re-searching-of}}
%%%%%%%%
In this section, we re-analyze the GWTC BBH events with component
masses lying between 10$M_{\varodot}$ and 100$M_{\varodot}$ available in the GWOSC \citep{abbott2023open,RICHABBOTT2021100658}. A general
idea behind conducting such a re-analysis is to test whether the eccentric
search can produce candidates in agreement with the earlier searches. This would enable us to test our capabilities for possibly contributing to conducting follow-up studies. We find that the eccentric search recovers a majority of the confident events with either three or a two detector coincidence without much fine-tuning. The FAR estimates here are limited by the duration ($\sim$ 4096 seconds) of the data analyzed over which the backgrounds are estimated. The backgrounds evaluated over longer durations of data are expected to provide more precise FAR estimates in general.

To illustrate this, in the case of \texttt{GW190521}, we further extend the search space for the component masses to 150$M_{\varodot}$ and use past 3.5 hours of data to compute the false alarm rate as shown in Fig.~\ref{fig:GW190521}. To tackle a larger search parameter space coverage, the swarm size is increased to about 2400 particles while the number of steps are kept fixed. The resulting search recovers the event with a FAR close to 1 per century. 

We have also included the event \texttt{GW170817} in Table~\ref{tab:gwtc_table} obtained with the eccentric BNS search as described in Section \ref{subsec:Eccentric-BNSs}. However, we have not considered any single detector events in our analyses, which we can address in the future.
\vspace{-0.6cm}
%%%%%%
\subsection{Runtime\label{subsec:Latency}}
%%%%%%%
Searches involving at least one neutron-star component are time-critical. To expedite the electromagnetic follow-up efforts, the LVK searches currently take a couple of seconds to few minutes to identify the event candidates. In the case of the template bank based searches, the process of matched-filtering can entirely proceed in parallel and the waveforms can be pre-computed and stored in the memory~\citep{nitz2018rapid}. However, the speed of the PSO based searches is limited by the cost of computing templates on the fly and the serial iterations. Thus, we try to minimize the time spent at each iteration with task parallelism. We split the set of CPU intensive computations in a given iteration across the available CPU cores. Each share of the total computation is tackled by an independent process. The processes exchange the required information for computing the fitness and for communicating the results using Message Passing Interface~(MPI). In other words, every core processes a fixed number of particles for all the iterations. Each iteration is synchronized and a \textit{master} process coordinates and drives the algorithm. The computations such as storing triggers to the disk for/or post-processing are also performed by the master process. Online analyses require that data be processed at least at the same rate as are recorded by the detectors. We have broadly timed the BNS search with 15 steps and about 5000 particles for the HL and the HLV networks as shown in Fig.~\ref{fig:Reduction-in-real-runtime}. We observe that the runtime is almost inversely proportional to the number of CPU cores available. 
\section{Conclusions}
We have explored a simple algorithm that allows us to search for
compact binary coalescences without using a pre-computed template bank. Instead, the PSO algorithm chooses template points on the fly while exploring an arbitrary dimensional search space. We exploit this feature to implement a five dimensional search to include the orbital eccentricity, in addition to the standard search parameters, namely the component masses and the aligned spins. If astrophysical sources carry a non-negligible eccentricity, the eccentric search improves the recovered signal-to-noise ratios. In this work, we investigate a search for BNS systems using around 75k matched-filtering operations whereas a significant part of the BBH parameter space can be tackled with about 30k matched-filtering operations. The algorithm utilizes these many number of operations to iteratively arrive at an optimal detection statistic and also provides reasonably unbiased point estimates for the source parameters like the chirp mass, the effective spin parameter and the eccentricity. We have described our eccentric search pipeline for multiple detectors and demonstrated it on many GWTC events.

We have presented a parallelization scheme to lower the real-runtime of the search by utilizing multiple CPU cores. In the case of an eccentric search, the speed and the accuracy of the available waveform families, is of a general concern. Some are fast but do not support large portions of the plausible parameter space, for example in the large mass ratio limit for binaries like the NSBHs, while others become slow while covering all the regions. Searches including the effects due to spin precession coupled with eccentricity is an interesting avenue whenever accurate waveform models for the complex physics come about. The mean anomaly parameter can also be incorporated into the analyses when suitable waveforms are available. Fortunately, waveform physics is an active area of research and development. With efficient waveforms spanning larger parts of the parameter space and some fine-tuning of the proposed method, it is possible to deploy an eccentric search that can be used in low-latency for the current generation of ground based detectors.
\begin{center}
\begin{figure}[t]
\includegraphics[scale=0.7]{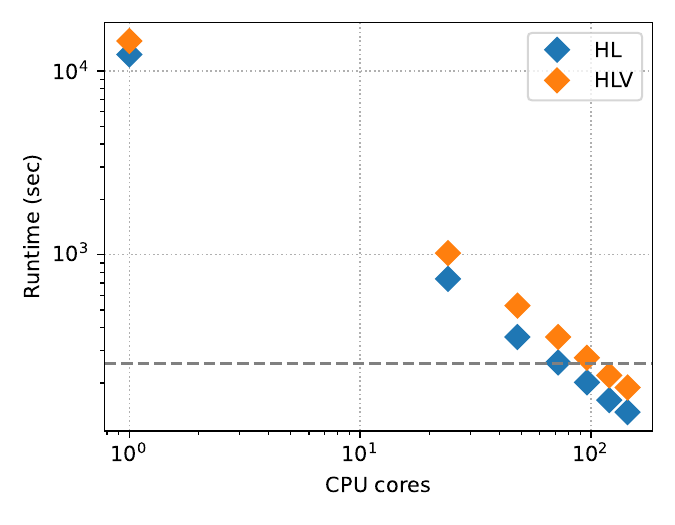}
\caption{Reduction in the runtime for a BNS search for $1M_{\varodot}-3M_{\varodot}$, aligned-spins between $\pm0.5$ and eccentricity upto 0.15 with multiple CPU cores suggests that the analysis can be performed in near realtime at a low computational budget. The horizontal dashed line indicates the duration of segment being processed. Note the runtimes lying below this line clear the fundamental criterion to be performed online.\label{fig:Reduction-in-real-runtime}}
\end{figure}
\par\end{center}
\vspace{-1cm}
\section{Acknowledgements}
This research has made use of data or software obtained from the Gravitational Wave Open Science Center (gwosc.org), a service of LIGO Laboratory, the LIGO Scientific Collaboration, the Virgo Collaboration, and KAGRA. LIGO Laboratory and Advanced LIGO are funded by the United States National Science Foundation (NSF) as well as the Science and Technology Facilities Council (STFC) of the United Kingdom, the Max-Planck-Society (MPS), and the State of Niedersachsen/Germany for support of the construction of Advanced LIGO and construction and operation of the GEO600 detector. Additional support for Advanced LIGO was provided by the Australian Research Council. Virgo is funded, through the European Gravitational Observatory (EGO), by the French Centre National de Recherche Scientifique (CNRS), the Italian Istituto Nazionale di Fisica Nucleare (INFN) and the Dutch Nikhef, with contributions by institutions from Belgium, Germany, Greece, Hungary, Ireland, Japan, Monaco, Poland, Portugal, Spain. KAGRA is supported by Ministry of Education, Culture, Sports, Science and Technology (MEXT), Japan Society for the Promotion of Science (JSPS) in Japan; National Research Foundation (NRF) and Ministry of Science and ICT (MSIT) in Korea; Academia Sinica (AS) and National Science and Technology Council (NSTC) in Taiwan.

Most of our analyses use the existing tools in the \texttt{PyCBC} software package. We fruitfully benefited from the discussions in the LVK Collaboration, the LIGO India Scientific Collaboration (LISC) and the PyCBC group. We acknowledge the use of the Sarathi computing cluster at IUCAA for computational/numerical work. This material is based upon work supported by NSF's LIGO Laboratory which is a major facility fully funded by the National Science Foundation. The article reserves the LIGO Document number LIGO-P2300164. The research was carried out at the Center of Excellence in Space Sciences, India (CESSI). CESSI is a multi-institutional Center of Excellence hosted by the Indian Institute of Science Education and Research (IISER) Kolkata and has been established through funding from the Ministry of Education, Government of India. Support also comes from the Council for Scientific and Industrial Research (CSIR), India through File No:09/921(0272)/2019-EMR-I. We thank the anonymous referee for providing the valuable feedback on the manuscript.
\bibliography{ecc_pso}
\bibliographystyle{apsrev4-1}
\end{document}